# Did you know that Economics is not only about money? The effect of popularisation talks on high school students' interest in the discipline*

Laura Padilla-Angulo[†], Diego Jorrat[‡], José-Ignacio Antón[§] and Javier Sierra[§]

[†] *Universidad Loyola Andalucía*

[‡] *University of Seville*

[§] *University of Salamanca*

**Abstract**

This paper evaluates the effect of a short, interactive popularisation talk on upper-secondary students' interest in Economics. This contrasts with previous research, which has primarily examined impersonal interventions to boost interest in Economics. The intervention presents Economics as an empirical social science engaged with real-world social problems. Using a cluster-randomised field experiment conducted during secondary-school campus visits in Spain, we find no statistically significant average effect on stated interest in studying Economics. However, the intervention generates substantial heterogeneity: those with stronger altruistic preferences become significantly more likely to express interest after the talk. These findings suggest that informational outreach may shape who perceives the discipline as aligned with their motivations, even if it does not substantially increase overall interest. More broadly, they indicate that presenting Economics as empirical and socially relevant may broaden the profile of those who consider the field.

## 1. Introduction

Economics plays a crucial role across a wide range of sectors, including private enterprise, public administration, consulting, health, the environment and finance. It underpins many public policy decisions and influences both grassroots initiatives and large-scale interventions (Megalokonomou et al., 2021). Economists apply data analysis and theoretical models to understand complex issues such as the effectiveness of public programmes, pandemic policy responses, wage disparities and behavioural dynamics with societal implications (American Economic Association [AEA], n.d.).

* Corresponding author: Laura Padilla-Angulo, Department of Economics, Universidad Loyola Andalucía, Campus Sevilla, Avda. de las Universidades s/n, 41704 Dos Hermanas, Sevilla (Spain), e-mail: lpadilla@uloyola.es. We received ethics approval from Universidad Loyola Andalucía in December 2022. We pre-registered the experimental design and analysis plan of this research prior to data collection on AsPredicted (#153,092, available at https://aspredicted.org/7e8ix.pdf) in November 2023. This research derives from the teaching innovation project "Science communicator for a day: talks on what Economics is and what economists do for Economics students". We thank P. Brañas, M. Christie, C. Figueroa-Sisniega, R. Muñoz de Bustillo, J. Ponce, P. Triunfo, M. Ramos and participants at the Southern Europe Experimental Team's Meeting (Seville, 2026) for helpful comments on a previous version of the paper. A. F. Soria Bollini provided excellent research assistance. Jorrat and Antón acknowledge financial support from the Spanish Ministry of Science, Innovation and Universities (PID2024-156629NB-I00 and PID2021-123875NB-I00, respectively).

Despite its societal importance, Economics has seen a steady decline in student enrolment across several countries in recent decades. In the U.S. and Australia, for instance, the number of Economics majors has decreased significantly (Avilova & Goldin, 2018; Dwyer, 2017; Livermore & Major, 2021; Lovicu, 2021; Marangos et al., 2013). In Europe, Eurostat (2025) data show that Economics graduates accounted for 2.1% of bachelor's degrees in 2015 and 1.6% in 2023. In contrast, Business and Administration—often perceived as a closely related discipline—had a much higher and growing share of graduates, rising from 15.9% in 2015 to 17.8% in 2023. In Spain, Economics graduates dropped from 2.5% of the total in 2001–2002 to 1.6% in 2023–2024, a decline that was more pronounced among women (from 2.1% to 1.1%) than men (from 3% to 2.3%) (Ministerio de Ciencia, Innovación y Universidades, 2025).

Beyond the shrinking numbers, diversity in Economics is scant (Bayer & Wilcox, 2019; Stansbury & Schultz, 2023). For example, Economics suffers from persistently low levels of gender and ethnic diversity. In the U.S., for example, Black, Hispanic, Native American and female students continue to be underrepresented in Economics programmes (Avilova & Goldin, 2018; Bayer & Rouse, 2016; Bayer & Wilcox, 2019; Pugatch & Schroeder, 2024). Some research indicates that students from these groups view the lack of diversity itself as a reason to avoid the field (Pugatch & Schroeder, 2024). This disparity extends to advanced degrees as well: women continue to make up only about 35% of Economics PhD students and around 30% of assistant professors globally—a figure that has stagnated since the mid-2000s (Lundberg & Stearns, 2019). Similarly, recent UK research reveals that Economics undergraduates are disproportionately from affluent and highly educated backgrounds, suggesting socio-economic exclusion as well (Paredes Fuentes et al., 2023). In addition, compared with other disciplines, Economics PhD recipients are more likely to come from highly educated families, and the gap has widened in recent decades (Stansbury & Schultz, 2023).

In Europe, gender gaps in Economics have not only persisted but also widened. According to Megalokonomou et al. (2022), between 2013 and 2018, only about 40% of Economics students were women. In Spain, the share of female Economics graduates dropped from nearly 51% in 2001–2002 to around 43% in 2023–2024 (Ministerio de Ciencia, Innovación y Universidades, 2025). This lack of diversity could limit the discipline's scope, as a homogenous academic environment may constrain the range of topics explored and hinder the emergence of innovative perspectives (Bayer & Rouse, 2016). Economics would benefit from greater gender and socio-economic diversity (Bayer et al., 2020b; Megalokonomou et al., 2021).

Research has identified a significant information gap as a key reason for both the declining interest in Economics and the persistent lack of diversity in the field. Inadequate information and misconceptions about what economists actually do seem to be at the root of these issues (Avilova & Goldin, 2018; Bayer et al., 2020a).

Many students are unaware of how Economics connects to sectors like health, education, inequality and development—areas typically not associated with the discipline (Avilova & Goldin, 2018; Bayer et al., 2020a; Lovicu, 2021). According to Bayer et al. (2020b), students often perceive a narrow and inaccurate career path in Economics, typically limited to roles in banking and finance, while remaining unaware of the broader policy and societal applications of the discipline.

This gap in understanding is especially pronounced among students from underrepresented or less privileged backgrounds, who often lack early exposure to professional pathways and may not receive guidance on the practical applications of different degrees. For many, college itself is an unfamiliar environment, and discussions about careers in fields like Economics may not occur within their families or communities.

Similarly, the literature documents distinct (less altruistic) behaviours among Economics students compared to those in other fields. Evidence on indoctrination effects is mixed, but the case for self-selection is strong (Miragaya-Casillas et al., 2023a, 2023b), yielding cohorts that are not fully representative of the broader student population.

To counter these misconceptions, the AEA and various researchers advocate for targeted information strategies to make Economics more accessible and appealing. Presenting Economics as an empirical and socially engaged discipline—rather than one narrowly focused on markets and finance—can help attract a wider, more diverse group of students (Bayer et al., 2019; Avilova & Goldin, 2018). Moreover, highlighting the relevance of Economics in areas that resonate more with traditionally underrepresented groups—such as behavioural sciences, health and development—may increase interest, especially among women.

Despite the growing recognition of these issues, relatively few studies have evaluated interventions aimed at making Economics more attractive, and most of the existing evidence is based on research in the U.S. (Bayer et al., 2020b; Livermore & Major, 2021). To help address this gap, this article examines whether popularisation talks—presenting Economics as an evidence-based social science focused on real-world problems—can increase interest in the field among high school students in Spain.

This study presents an empirical test of whether providing high school students with information about what economists do and the types of problems they address through popularisation talks can increase their interest in pursuing undergraduate studies in Economics. Such an impact could vary by socio-economic status and distributional preferences, thereby affecting the gender and social diversity of the Economics student pool.

Our findings indicate that, while the intervention does not yield a statistically significant increase in average interest in Economics, it has economically meaningful effects and gives rise to substantial heterogeneity. In particular, students with stronger altruistic preferences become significantly more likely to report interest in studying Economics after the talk. This is a promising result, particularly considering that university student populations in Spain are generally less diverse than secondary school cohorts (Sierra et al., 2025). Thus, such interventions might play a role in shaping the composition—and potentially expanding the diversity—of future Economics students.

This research contributes to the literature in several ways: by examining a personal and interactive intervention (as opposed to the impersonal communication methods typically studied, such as emails), by targeting high school students rather than university undergraduates, by assessing the role of distributional preferences in shaping responses to informational interventions and by focusing on a European context—specifically Spain, where students must choose their field of study before entering university. This last feature contrasts with the U.S. system and may have important implications for when and how interventions are most effective. The study was pre-registered before data collection, and the hypotheses, primary outcome and baseline specifications were laid out ex ante in a pre-analysis plan. Any additional analyses are clearly labelled as robustness checks.

The remainder of the article proceeds as follows. Section 2 reviews related literature; Section 3 details the field experiment, intervention and data; Section 4 reports the main results; Section 5 discusses their implications; and Section 6 concludes with directions for future research.

## 2. Literature review and hypothesis development

### 2.1. The importance of diversity in Economics

The underrepresentation of diverse groups in Economics has been widely documented (Avilova & Goldin, 2018; Bayer & Rouse, 2016; Bayer & Wilcox, 2019; Castellanos Rodríguez, 2021; Livermore & Major, 2021; Lovicu, 2021; Megalokonomou et al., 2022; Miragaya-Casillas et al., 2023a, 2023b). This lack of diversity within the student body and, by extension, the profession poses significant challenges to the field. It limits the scope of research questions addressed by economists and can result in narrow or biased policy recommendations due to the absence of varied perspectives and lived experiences (May et al., 2014; Pugatch & Schroeder, 2024; Bayer et al., 2020a).

For example, Lundberg and Stearns (2019) show that female Ph.D. recipients in Economics are more likely than their male counterparts to focus on labour and public economics and less likely to specialise in macroeconomics and finance. Women also tend to gravitate towards applied areas of the discipline. Similarly, May et al. (2014) find that female economists are generally more supportive of government intervention than male economists, who tend to prefer market-based solutions.

Given the influential role economists play in shaping public policy, the presence of diverse voices becomes crucial. As Mester (2021) argues, policy decisions are more likely to reflect the needs of a broad population when those making or informing these decisions represent a wide range of backgrounds and experiences. In her speech at the Allied Social Science Associations Annual Meeting, organised by the AEA, Mester emphasised that diverse groups process information more effectively because individuals must engage with and challenge alternative viewpoints.

Another key concern is the shortage of visible role models for underrepresented groups—such as Black, Latinx and Native American students—in Economics. Bayer et al. (2020b) underscore the significance of role models as a critical factor influencing minority students' decisions to pursue a career in the field. The absence of representation not only hinders student engagement but also contributes to persistent disparities in Economics education and careers.

### 2.2. The role of information provision in addressing the diversity gap

A growing body of research suggests that one of the primary reasons for the persistent lack of diversity in Economics is a widespread information gap regarding what the discipline entails and its relevance to a broad range of sectors, including health, education, inequality, development and other pressing social issues (Avilova & Goldin, 2018; Bayer et al., 2020a; Lovicu, 2021).

Bayer et al. (2020b) highlight that many undergraduate students lack a clear understanding of the career pathways available to Economics graduates. There is a common perception that the degree mainly leads to jobs in banking, while its connections to public policy and roles in both public and private institutions are often overlooked. In contrast, career paths following a business degree tend to be easier for students to understand. This lack of clarity is particularly acute among students from underrepresented backgrounds.

A key insight from Bayer et al. (2020b) is that many students from underrepresented backgrounds enter higher education with limited exposure to professional careers, including those in Economics. For some, especially first-generation college students or those from working-class families, college itself can feel unfamiliar and abstract, and information about potential career paths is often scarce. This lack of early guidance and exposure can contribute to lower awareness of the varied opportunities an Economics degree can offer.

To address this challenge, several researchers advocate for making the scope and relevance of Economics more visible and accessible to a broader audience. Bayer et al. (2019), for instance, emphasise that informing students about the diversity of issues that economists study, as well as the different career trajectories available, can help attract individuals with a wide range of goals and backgrounds. Doing so could increase participation among students who might otherwise not have considered Economics a viable or relevant field of study.

Furthermore, informal influences—such as advice from family and friends—often promote Economics to male students more than to female students, typically by emphasising its connection to finance and banking (Avilova & Goldin, 2018). Conversely, highlighting aspects of Economics that overlap with psychology and human behaviour may be more effective in attracting women, particularly given the popularity of behavioural sciences among female students (Mester, 2021).

As mentioned above, a growing body of evidence suggests that the "economists are less altruistic" pattern is largely driven by self-selection, as Economics students already display more self-regarding social preferences at entry. One plausible mechanism behind this sorting is information and identity: prospective students may hold biased or incomplete beliefs about what Economics is (for example, that it is primarily about self-interest, competition and "defection" in strategic settings) and may also anticipate social norms associated with the economist stereotype. If so, students with more individualistic preferences may perceive a better "fit" and select into Economics, while more prosocial students may avoid it, reinforcing compositional differences even before any coursework effects. This hypothesis is consistent with work that explicitly links the economist stereotype and identity to cooperative behaviour in strategic interactions (Lanteri & Rizzello, 2014). More generally, there is strong causal evidence that students' beliefs about major-specific characteristics are often inaccurate and that they update these beliefs when provided with information, with implications for (stated) major choice (Wiswall & Zafar, 2015). Together, these findings motivate the prediction that correcting misconceptions about Economics (content, norms, social impact and typical student profiles) could shift the composition of entrants, partly attenuating the observed self-selection in prosocial preferences.

### 2.3. Interventions providing information on Economics

Given the widespread lack of understanding of what Economics entails, several researchers have developed interventions to address this information gap by providing insights into various aspects of the discipline, including research topics, career opportunities and potential earnings. Some studies also evaluate the effectiveness of these interventions in increasing students' interest in Economics and, consequently, promoting diversity within the field.

For instance, Bayer et al. (2019) conducted a randomised controlled trial (RCT) involving 2,710 students across nine U.S. colleges. The study tested whether informational "nudge" emails from faculty members could influence the number of Economics courses which incoming university students take. One email simply welcomed students and encouraged them to consider enrolling in Economics. The second email included additional content highlighting the breadth of research topics in Economics and provided links to resources on the AEA website, such as examples of economists' work and career testimonials. Both emails increased enrolment in Economics courses, but the second—richer in content—had a significantly greater effect, suggesting that how Economics is presented plays a critical role in shaping students' interest. This effect seems to be larger among first-generation students.

Pugatch and Schroeder (2024) similarly used an RCT with over 2,200 students enrolled in Principles of Economics to assess how different types of informational nudges affected the socio-economic and ethnic diversity of students majoring in Economics. Four variations of messages were tested: (1) a basic message encouraging enrolment based on the department's standard description; (2) the same message with added information on graduates' earnings at different career stages; (3) a message with a video from the AEA highlighting career opportunities in Economics; and (4) a message featuring a video with diverse current and former Economics students. The interventions increased the proportion of first-generation, migrant and underrepresented minority students declaring an Economics major—but notably, only among male students.

Further research has investigated the gendered effects of such interventions. Pugatch and Schroeder (2021) explored, in a related RCT involving over 2,000 students, whether responses to these messages differed by gender. Students were randomly assigned to receive one of the four previously mentioned messages or no message at all (control group). In a second phase, students with a grade of B− or higher received follow-up messages encouraging them to major in Economics. Some of these follow-ups also addressed students who had underperformed relative to expectations, aiming to boost their confidence. The study found that while the basic message increased the share of male students majoring in Economics by two percentage points, it had no measurable effect on female students, indicating that women may require more interactive or personalised engagement.

In line with this, Li (2018) implemented an RCT with 450 students in Introductory Economics at a large U.S. university. Female students with above-median grades received an email recognising their achievement and encouraging them to consider majoring in Economics. The intervention included a video on career prospects and salary expectations, a pamphlet and information on grade distribution. When combined with mentoring, this significantly increased the likelihood that these female students would choose Economics as a major.

Porter and Serra (2020) also found that in-person engagement can be more effective for female students. In their RCT with 627 female students, a classroom visit by a female role model increased the probability that students would major in Economics. In contrast, studies using less personal approaches—such as Antman et al. (2025), who sent grade-related emails, or Halim et al. (2022), who used emails and physical letters to provide information and encouragement—did not find significant effects among female students.

Bedard et al. (2021) examined similar nudging strategies with 2,338 students who had received at least a C in an introductory microeconomics course. Faculty members sent personalised letters with information about Economics majors, potential careers and an invitation to an informational meeting. A subset of students with a B or higher received an additional message acknowledging their academic achievement and encouraging them to pursue Economics. These personalised messages increased both attendance at the meeting and the likelihood of majoring in Economics, with the most substantial effects observed among Hispanic students, particularly women.

### 2.4. Hypotheses

Guided by this literature, and as stated in our pre-analysis plan, we formulate the following hypotheses:

- H1: Popularisation talks can increase interest in Economics among high school students.

In line with previous findings, we also expect the treatment effect to vary by gender, socio-economic background and the proxy for social preferences. In particular:

- H2: The effect of providing information is larger among female students.
- H3: The effect of providing information is larger among students whose parents have low educational attainment (used as a proxy for lower socio-economic status).

— H4: The effect of providing information is larger among students with stronger distributional preferences (i.e. those who are more averse to unequal outcomes). This hypothesis implies an interaction effect between the treatment and the aversion to inequality. We predict a positive multiplicative effect, although this hypothesis is exploratory and lacks prior empirical evidence.

## 3. Methodology

### 3.1. Field experiment

We conducted a pre-registered field experiment between November 2023 and February 2024, in which high school students were exposed to a 15-minute popularisation talk about what economists do in their jobs and the types of problems they address. Our setup involved visits by students in the last year of secondary education (the second year of the Baccalaureate) to a Spanish private university. The university organises this activity annually in collaboration with various high schools in the region where it is located (Andalusia, Spain).

The visit enables prospective university students to become familiar with the institution's infrastructure and academic offerings (i.e., the undergraduate degrees available for study). The visit includes an informative talk about the undergraduate degrees offered by the institution, organised by the university's Student Orientation and Information Services. From the university's perspective, the primary objective of the activity is to attract new students for the upcoming academic year. Organisers ask students attending these informative talks to complete a survey form where, along with their personal data and some academic details (school, year, Baccalaureate track and average mark), they have to indicate three degrees offered by the university (in descending order of preference) on which they would like to receive further information (indicating it with a number), whether they are interested in a degree that is not in the list of programmes available at the university and whether they plan to study at the institution (if so, in which of the three campuses of the university in the region). In our study, we asked the students about the three majors they are most interested in (in decreasing order). The question was part of a survey form that students were required to fill out during visits to the university, where they received information on the institution's academic offerings. It is worded as follows: "Among the degrees offered by the university, indicate, with the number of the degree, which would be your first (most preferred), second and third choice?".

Randomisation was conducted at the school level prior to the visits. The university's Student Orientation and Information Service provided a list of high schools that typically participate in the annual visit programme. We then randomly assigned schools to the treatment or control condition using a simple Bernoulli-type randomisation procedure. We assigned 12 schools to the treatment condition (538 students) and 14 schools to the control group (775 students).[1] In both cases, students attended the regular orientation talk provided annually by the university's student office, which offers general information on the institution's academic programmes. In the treatment schools, this talk was immediately followed by an additional 15-minute session on what economists actually do, whereas students in the control schools received only the standard orientation talk. At the end of each session, all students completed the same survey administered by the university's Student Orientation and Information Services. All talks were delivered by the same presenter—an associate professor at the host institution.

Furthermore, the schools visited the campus on different days and/or at different times, and orientation activities were organised separately by each visiting group. This arrangement eliminates the possibility of interactions across schools and the likelihood of spillovers between treatment and control groups.

The popularisation talk (treatment) emphasises two propositions that are widely shared within the profession but often unfamiliar to students—and that prior research suggests can increase interest in studying Economics (Bayer et al., 2020a): (i) Economics is a social science concerned with solving social problems and improving welfare and (ii) modern Economics is fundamentally empirical, using the scientific method and quasi-experimental approaches to study causal relationships. The presentation illustrates these points with real-world examples drawn from recent, influential economics papers. Appendix A includes both the presentation used in the popularisation talks and the survey form.

The study protocol received approval from the Ethics Committee of Universidad Loyola Andalucía. All questionnaires were collected anonymously, and no personally identifiable information was recorded. Participants were high school students aged 15 or older and completed the survey voluntarily during the campus visit; therefore, parental consent was not required under the applicable ethical guidelines.

### 3.2. Measures

Our pre-specified dependent variable captures interest in pursuing an undergraduate degree in Economics. It is a binary indicator equal to one if Economics, including double degrees, is listed among a student's top three degree preferences and zero otherwise.

[1] The pre-analysis plan anticipated a larger sample of schools and students. In practice, participation in the visit programme during the study period resulted in a smaller realised sample than initially envisaged in the pre-analysis plan.

Regarding socio-economic status, we indirectly measure it using the answer to the question: "Have either of your parents completed university studies?". We construct a dummy that takes the value of one if the answer is yes and zero if the answer is no.

Following Fehr and Schmidt (1999) and adapting the question designed by Brañas-Garza et al. (2022), we ask the students to indicate which option they agree with most: (A) "I am not worried about how much money I have, what worries me is that there are people who have more money than I have."; (B) "I am not worried about how much money I have, what worries me is that there are people who have less money than I have." Option B represents a measure of compassion, reflecting advantageous aversion to inequality. Accordingly, we define *altruistic* as a binary variable equal to 1 if the student selected option B, capturing compassion or advantageous inequality aversion.

### 3.3. Empirical strategy

To test H1, we estimate the following linear probability model as set out in the pre-analysis plan:[2]

$$Y_{is} = \alpha + \beta T_s + \gamma X_{is} + \varepsilon_{is} \quad (1)$$

where $Y_{is}$ is the outcome for student $i$ in school $s$, $Ts$ indicates whether the school was assigned to the treatment condition (and 0 otherwise), $X_{is}$ is a vector of student-level controls and $\varepsilon_{is}$ is a random disturbance. This vector includes gender, parental education, distributional preferences, whether the student is in the final year of high school and whether the school is public or private. Since randomisation was conducted at the school level, as planned, standard errors are clustered at the school level (Abadie et al., 2023). Given the small number of clusters (26 schools, 12 treated), we report conventional cluster-robust standard errors and the associated inference, alongside $p$-values from wild cluster bootstrap (WCB) procedures based on 9,999 replications and Rademacher weights. WCB improves size control in settings with few clusters and often yields more conservative inference than conventional cluster-robust methods (MacKinnon & Webb, 2018). $\beta$ is our parameter of interest and captures the average treatment effect of being exposed to the additional 15-minute popularisation talk.

We also examine whether the treatment effect varies by student characteristics, as laid out in the pre-analysis plan, in order to test hypotheses H2, H3 and H4. To do so, we interact the treatment indicator with gender, parental education as a proxy for socio-economic status and distributional preferences (inequality aversion). As additional robustness checks, not pre-specified in the pre-analysis plan, we also estimate *logit* models and use multiple imputation for missing covariates; the results are qualitatively similar.[3]

[2] In the pre-analysis plan, the empirical specification inadvertently featured a group (school) fixed effect. Because assignment is at the school level, such fixed effects are collinear with the treatment indicator and would preclude identification of the average treatment effect. We therefore omit this term in all analyses.

### 3.4. Data and balance check

The final dataset includes 1,313 students from 26 high schools participating in the study. Table 1 reports descriptive statistics for the main variables. The main outcome, interest in studying Economics, equals 1 for 7% of students. Regarding key covariates, 61% of respondents are classified as altruistic, 53% are female, and 82% report that at least one of their parents holds a university degree. About 88% of students are in the second year of upper-secondary school. In terms of school type, 41% attend private schools, 41% charter schools and the remaining 18% public schools. The average grade point average (GPA), self-reported by students, is 8.5 (out of 10).

Table 1. Summary statistics for the main variables

| | No. of observations | Mean | Standard deviation | Minimum | Maximum |
|---|---|---|---|---|---|
| Interest in Economics | 1,313 | 0.07 | | 0.00 | 1.00 |
| Female | 1,305 | 0.53 | | 0.00 | 1.00 |
| Parents with a university degree | 1,219 | 0.82 | | 0.00 | 1.00 |
| Altruistic | 895 | 0.61 | | 0.00 | 1.00 |
| GPA | 1,069 | 8.55 | 1.20 | 0.00 | 10.00 |
| Senior | 1,313 | 0.88 | | 0.00 | 1.00 |
| Charter school | 1,313 | 0.41 | | 0.00 | 1.00 |
| Public school | 1,313 | 0.17 | | 0.00 | 1.00 |

Before analysing the main results, we first examine whether the treatment and control groups are balanced in terms of observable characteristics. Table 2 reports detailed balance tests for each baseline covariate. Using conventional cluster-robust inference at the school level, parental university education and public-school status show statistically significant differences at the 10% level. However, when using WCB inference, only parental university education remains statistically significant at this level.

The magnitude of the differences is modest. The imbalance may stem from the uneven allocation of public schools across arms (four in treatment versus one in control), which could mechanically affect the composition of parental education. Following Kerwin et al. (2024), we assess overall covariate balance using a joint orthogonality test with randomization inference. Excluding variables with higher rates of missing data (altruistic and GPA), the test yields an $F$-statistic of 1.95 and a $p$-value of 0.485, confirming that random assignment achieved satisfactory overall balance between treatment and control schools.

[3] In these robustness checks, we report cluster-robust standard errors. For the *logit* specifications, inference for average marginal effects is based on delta-method standard errors, while multiple-imputation estimates are pooled using Rubin's rules. WCB-*t* refinements are mainly used for coefficient-level inference in linear models with few clusters. Extending analogous refinements to marginal effects or pooled multiple-imputation estimates would require a separate resampling approach.

As mentioned, almost 32% of the observations are missing for the variable altruistic, 19% for GPA and 42% when considering missing values in either variable. To assess whether this non-response could bias the results or compromise the experiment's internal validity, we perform a series of complementary tests.

Table B1 in Appendix B reports the results of the attrition regressions based on three non-response indicators: one for altruistic, one for GPA and one combining both. Panel A tests for differential attrition by regressing each non-response dummy on treatment status using WCB inference at the school level. In all three cases, the treatment coefficient is statistically insignificant ($p$-values are 0.581 for altruistic, 0.115 for GPA, and 0.496 for the combined indicator), indicating no evidence of systematic differential attrition across groups.

Panel B extends the analysis by adding baseline covariates to assess attrition conditional on observables. The treatment coefficient remains statistically insignificant in all specifications. Moreover, the joint $F$-tests fail to reject the null hypothesis that all covariate coefficients are equal to zero (WCB $p$-values of 0.232, 0.206, and 0.245). Taken together, these results provide no evidence that attrition is systematically related to treatment assignment or observable characteristics, suggesting that missingness is unlikely to threaten the study's internal validity.

Finally, Table B2 in Appendix B examines whether randomization remains balanced within the non-missing samples. The table replicates the balance tests reported in Table 2 but restricts the sample to observations with non-missing data for altruistic preferences (Panel A), GPA (Panel B) and both variables jointly (Panel C). The pattern of differences between treatment and control groups closely mirrors that observed in Table 2: a small number of covariates—notably parental university education and public-school status—display modest differences across groups. To assess overall balance within each restricted subsample, we implement a joint orthogonality test based on randomization inference. Across all panels, the tests fail to detect systematic imbalance, indicating that attrition has not materially altered the original random assignment.

Table 2. Balance across treatment and control groups

| | (I) Control | (II) Treatment | (III) Difference (II) – (I) |
|---|---|---|---|
| Female | 0.548 | 0.497 | −0.051 |
| | (0.498) | (0.500) | (0.053) |
| | | | [0.356] |
| Parents with a university degree | 0.869 | 0.745 | −0.124* |
| | (0.337) | (0.436) | (0.070) |
| | | | [0.074] |
| Altruistic | 0.599 | 0.625 | 0.026 |
| | (0.491) | (0.485) | (0.040) |
| | | | [0.554] |
| GPA | 8.598 | 8.463 | −0.135 |
| | (1.204) | (1.203) | (0.250) |
| | | | [0.650] |
| Senior | 0.888 | 0.861 | −0.027 |
| | (0.316) | (0.347) | (0.117) |
| | | | [0.812] |
| Charter school | 0.465 | 0.329 | −0.136 |
| | (0.499) | (0.470) | (0.224) |
| | | | [0.586] |
| Public school | 0.041 | 0.346 | 0.304* |
| | (0.199) | (0.476) | (0.158) |
| | | | [0.106] |
| Joint orthogonality test | | | |
| *F*-statistic | $F(5, 25) = 1.950$ | | |
| *p*-value | 0.485 | | |
| No. of observations | 775 | 538 | |

*Notes:* Columns (I) and (II) report mean values for the control and treatment groups, respectively, with standard deviations in parentheses. Column (III) reports the difference in means between the treatment and control groups, estimated from OLS regressions of each covariate on the treatment indicator. Cluster-robust standard errors for the estimated differences are reported in parentheses and WCB *p*-values in brackets. Asterisks in column (III) denote statistical significance of the difference in means based on conventional cluster-robust standard errors: *** $p < 0.01$, ** $p < 0.05$, * $p < 0.10$. The joint orthogonality test follows Kerwin et al. (2024) and is based on randomisation inference. WCB is not used for this joint test, as the procedure relies on the exact random assignment mechanism rather than asymptotic approximations and remains valid with a limited number of clusters.

Taken together, the evidence from Tables B1 and B2 indicates that attrition does not depend on treatment assignment and that the random allocation of schools remains valid within the non-missing samples. There is no systematic evidence that non-response is jointly explained by observable characteristics, suggesting that attrition is unlikely to compromise the experiment's internal validity.

## 4. Results

In this section, we present the pre-specified estimates of the average treatment effect of the popularisation talk on our outcome (interest in studying Economics) and the planned heterogeneity analyses to test the different hypotheses. We then report additional robustness checks.

### 4.1. Average effect of the popularisation talk

Table 3 reports the estimated effect of the popularisation talk on students' interest in studying Economics. Column (I) presents a simple regression of the outcome on the treatment indicator, while Column (II) adds individual- and school-level covariates. Standard errors are clustered at the school level and, given the limited number of clusters (26 schools), WCB *p*-values are also reported.

Across specifications, point estimates are small and statistically insignificant under both conventional cluster-robust inference and WCB inference. In Column (I), the coefficient is 0.022 (WCB *p*-value = 0.427), suggesting that the popularisation talk increased students' reported interest in Economics by about 2.2 percentage points. Given that only 7.5% of students in the control group reported interest, this represents a relative increase of approximately 29%. When individual and school controls are included (Column II), the coefficient remains positive and similar in magnitude (0.021, WCB *p*-value = 0.556). Column (III) reports the results obtained under the same specification, using the restricted sample for which altruism and GPA are observed. In this subsample, the estimated coefficient is almost zero (0.003, WCB *p*-value = 0.930). This attenuation appears to reflect differences in sample composition rather than the inclusion of additional covariates, as the attrition analyses in Tables B1 and B2 show that non-response in GPA and altruism is unrelated to treatment, while the restricted samples display modest differences in observable characteristics such as parental education and school type.

Overall, these results indicate that the popularisation talk did not significantly increase students' stated interest in Economics, although the estimated magnitudes in the first two specifications are economically meaningful and align with the direction of effects found in comparable informational interventions. For example, Bayer et al. (2019) found that their most comprehensive treatment, consisting of a series of emails that involved a welcome, an encouragement to take a course from the school's economics department and information highlighting the diversity of research and researchers in the discipline, increased the likelihood of completing an economics course by 3 percentage points, roughly 20% of the percentage of students taking a similar course in the control group.

Table 3. Average treatment effects of the popularisation talk on students' interest in Economics (OLS estimates)

| | (I) | (II) | (III) |
|---|---|---|---|
| Treatment | 0.022 | 0.021 | 0.003 |
| | (0.025) | (0.031) | (0.028) |
| | [0.427] | [0.556] | [0.930] |
| Female | | −0.059** | −0.050* |
| | | (0.021) | (0.024) |
| Parents with a university degree | | −0.028 | −0.022 |
| | | (0.019) | (0.030) |
| Senior | | 0.024 | 0.005 |
| | | (0.018) | (0.038) |
| Charter school | | 0.017 | 0.002 |
| | | (0.023) | (0.023) |
| Public school | | 0.006 | 0.020 |
| | | (0.036) | (0.041) |
| GPA | | | 0.001 |
| | | | (0.008) |
| Altruistic | | | −0.001 |
| | | | (0.020) |
| No. of observations | 1,313 | 1,211 | 730 |
| $R^2$ | 0.002 | 0.018 | 0.014 |

*Notes:* Each column reports OLS estimates of the effect of the popularisation talk on students' interest in studying Economics under different sets of control variables. All specifications include an intercept. Standard errors clustered at the school level (26 clusters) are reported in parentheses. WCB *p*-values are reported in brackets to account for the small number of clusters. Asterisks denote statistical significance based on cluster-robust standard errors: *** $p < 0.01$, ** $p < 0.05$, * $p < 0.10$.

At the same time, the experiment was not powered to detect modest average treatment effects with much precision. Based on the standard error of the treatment coefficient in Table 3, the minimum detectable effect at conventional significance levels and 80% power is approximately 7 to 9 percentage points, depending on the specification. This is substantially larger than the estimated average effect of about 2 percentage points. The lack of statistical significance for the average treatment effect should therefore be interpreted as reflecting limited precision in a cluster-randomised design with a relatively small number of schools, rather than as evidence of no effect.

### 4.2. Heterogeneous effects

Following our pre-registered analysis plan, we estimate interaction models to assess heterogeneous treatment effects by gender (H2), parental education (H3) and altruism (H4), using the same specification as in the baseline analysis. Table 4 summarises the results. In these models, the treatment coefficient captures the effect of the talk for the reference group, while the interaction term measures the differential effect for the subgroup of interest. The combined effect (which is the sum of the treatment and interaction coefficients) gives the total effect for the other subgroup (with its clustered standard error and its statistical significance) and is reported in the last two rows.

Panel A assesses the interaction with gender (H2). The coefficient on treatment is small (between -0.002 and -0.025) and statistically insignificant across all specifications (WCB *p*-values between 0.611 and 0.952). The interaction term between treatment and female is positive (between 0.039 and 0.055) but not statistically significant (WCB *p*-values between 0.242 and 0.308), and the combined effect for female students (the sum of the treatment main effect and the interaction between treatment and female) is also small (between 0.033 and 0.043) and insignificant (WCB *p*-values between 0.169 and 0.279). Overall, we find no evidence that the treatment differentially affects male and female students.

Panel B examines the effect of heterogeneity by parental education (H3), using the highest parental attainment as a proxy for socio-economic background. Among students whose parents do not hold a university degree, the estimated treatment effect is negative but statistically insignificant, ranging from -0.027 to -0.040 (WCB *p*-values between 0.472 and 0.752). For students with university-educated parents (the interaction term), the coefficient is positive and not significantly different from zero across all specifications (coefficients between 0.052 and 0.066; WCB *p*-values between 0.160 and 0.608). The combined effect for students with university-educated parents is small (0.012–0.036) and statistically insignificant (WCB *p*-values between 0.285 and 0.627). Overall, we find no robust evidence that the treatment effect varies by parental education.

Regarding distributional preferences (Panel C), corresponding to H4, the estimated treatment effect for non-altruistic students (the reference group) is negative across all specifications (−0.054 to −0.076) and statistically significant at the 10% level in the fully controlled specification (WCB *p*-value = 0.070). The interaction term for altruistic students is positive (between 0.119 and 0.129) and statistically significant in all specifications (WCB *p*- values between 0.001 and 0.002), indicating that the treatment effect is more positive for altruistic than for non-altruistic students.

In addition, the combined effect for altruistic students—i.e., the sum of the treatment and interaction coefficients—ranges from 0.053 to 0.065 and is statistically significant across specifications (WCB *p*-values between 0.011 and 0.095). Taken together, these results support H4: the popularisation talk increases interest in studying Economics among altruistic students.

### 4.3. Robustness checks

The analyses in this subsection are additional robustness checks and were not pre-specified in the pre-analysis plan. To assess the stability of our results, we re-estimate the specifications using *logit* models and report average marginal effects to ensure comparability with the baseline linear probability models. Standard errors are clustered at the school level. We present these supplementary results in Tables B3, B4 and B5 in Appendix B.

Table B3 reports the *logit* estimates for the baseline specifications. The estimated effect of the talk on interest in Economics is small and statistically insignificant across models and remains unchanged after the inclusion of different controls. The sign and magnitude of all coefficients closely mirror those obtained from Table 3. Table B4 examines treatment effect heterogeneity using *logit* models with interactions for gender (Panel A), parental education (Panel B) and altruistic preferences (Panel C). Group-specific treatment effects are computed using post-estimation marginal effects evaluated at different values of the interacting variable. Panels A and B show no evidence of heterogeneous effects by gender or parental education. Finally, Panel C replicates and strengthens the pattern observed in Table 4. The treatment effect is negative for non-altruistic students, whereas it is positive and statistically significant for altruistic students across all specifications, indicating a higher probability of reporting interest in Economics among altruistic individuals. These results provide robust support for H4, confirming that the heterogeneous effect of altruistic preferences is not sensitive to the choice of functional form.

Table 4. Effect heterogeneity by gender, parental education and distributional preferences (OLS estimates)

| | (I) | (II) | (III) | (IV) | (V) | (VI) | (VII) | (VIII) | (IX) |
|---|---|---|---|---|---|---|---|---|---|
| | | Panel A | | | Panel B | | | Panel C | |
| | Heterogeneity by gender (H2) | | | Heterogeneity by parental education (H3) | | | Heterogeneity by distributional preferences (H4) | | |
| Treatment | −0.002 | −0.004 | −0.025 | −0.030 | −0.027 | −0.040 | −0.054 | −0.057 | −0.076* |
| | (0.037) | (0.041) | (0.044) | (0.038) | (0.047) | (0.076) | (0.032) | (0.040) | (0.037) |
| | [0.952] | [0.933] | [0.611] | [0.472] | [0.684] | [0.752] | [0.115] | [0.220] | [0.070] |
| Female | −0.073** | −0.078** | −0.072** | | −0.057** | −0.048* | | −0.055** | −0.051** |
| | (0.032) | (0.032) | (0.033) | | (0.021) | (0.024) | | (0.023) | (0.025) |
| Parents with a university degree | | −0.026 | −0.020 | −0.061** | −0.058** | −0.047 | | −0.012 | −0.022 |
| | | (0.019) | (0.030) | (0.029) | (0.027) | (0.054) | | (0.022) | (0.031) |
| Altruistic | | | −0.002 | | | −0.001 | −0.064** | −0.052** | −0.050* |
| | | | (0.020) | | | (0.020) | (0.024) | (0.024) | (0.026) |
| Treatment × female | 0.040 | 0.047 | 0.057 | | | | | | |
| | (0.037) | (0.038) | (0.047) | | | | | | |
| | [0.308] | [0.253] | [0.242] | | | | | | |
| Treatment × parents with a university degree | | | | 0.066 | 0.058 | 0.052 | | | |
| | | | | (0.044) | (0.044) | (0.070) | | | |
| | | | | [0.160] | [0.277] | [0.608] | | | |
| Treatment × altruistic | | | | | | | 0.119*** | 0.121*** | 0.129*** |
| | | | | | | | (0.031) | (0.031) | (0.032) |
| | | | | | | | [0.001] | [0.002] | [0.002] |
| No. of observations | 1,305 | 1,211 | 730 | 1,219 | 1,211 | 730 | 895 | 860 | 730 |
| $R^2$ | 0.014 | 0.020 | 0.018 | 0.006 | 0.019 | 0.016 | 0.016 | 0.030 | 0.031 |
| Treatment + interaction | 0.038 | 0.043 | 0.033 | 0.036 | 0.030 | 0.012 | 0.066** | 0.065* | 0.053* |
| | (0.025) | (0.0314) | (0.026) | (0.029) | (0.031) | (0.023) | (0.025) | (0.033) | (0.026) |
| | [0.169] | [0.279] | [0.220] | [0.285] | [0.416] | [0.627] | [0.011] | [0.095] | [0.081] |

*Notes:* Each column reports OLS estimates of the effect of the popularisation talk on students' interest in studying Economics, including the corresponding interaction term, under different sets of control variables. All specifications include an intercept; specifications in columns (I), (IV) and (VII) include no additional controls; columns (II), (V) and (VIII) add type of school and senior status; columns (III), (VI) and (IX) additionally include GPA. The row labelled *treatment + interaction* reports the total effect for the interacted category. Standard errors clustered at the school level (26 clusters) are reported in parentheses. WCB *p*-values are reported in brackets to account for the small number of clusters. Asterisks denote statistical significance based on cluster-robust standard errors: *** $p < 0.01$, ** $p < 0.05$, * $p < 0.10$.

As a final robustness check, we address the loss of observations due to missing data in two baseline covariates—altruistic behaviour and GPA—thereby reducing the estimation sample in our most saturated specifications. To address potential concerns related to sample size and composition, we use multiple imputation by chained equations (Royston & White, 2011), generating 50 imputations with 10 burn-in iterations. Altruistic behaviour is imputed using a *logit* model, whereas GPA is imputed using linear regression, reflecting the binary and continuous nature of these variables, respectively. The imputation model includes baseline covariates (gender, parental university education, upper-secondary grade level and school type) and school fixed effects. We then re-estimate both the baseline specifications and the interaction models for gender, parental education and altruistic behaviour using the imputed datasets. Standard errors in all post-imputation regressions are clustered at the school level. The results (Table B5) are qualitatively consistent with our earlier findings: the average treatment effect is statistically insignificant, while the interaction between the treatment and altruistic behaviour remains positive and statistically significant, reinforcing the evidence in support of H4 and indicating that the main conclusions are not driven by missing data or sample attrition.

## 5. Discussion of results

The objective of this study is to assess whether a short, interactive, in-person popularisation talk about what economists do can increase high school students' interest in studying Economics. Our results suggest no significant effect of the popularisation talk. However, the point estimates imply a sizeable increase—approximately 29% relative to the baseline share of students reporting interest in Economics—indicating that even brief informational interventions may meaningfully shift stated preferences, albeit with limited statistical precision. These findings motivate further studies with larger samples to achieve greater statistical power.

Our results align with a growing body of literature showing that perceptions and information about Economics play a crucial role in shaping educational choices. Bayer and Rouse (2016) argue that Economics is often perceived as narrowly focused on markets and finance, rather than as a discipline concerned with social issues and public policy. A series of recent interventions has demonstrated that providing information, reframing course content or altering the presentation of Economics can positively impact students' interest and participation, particularly among underrepresented groups (Bayer et al., 2019; Bayer et al., 2020a; Bedard et al., 2021).

A key distinguishing feature of our work is that our design enables us to take a step further by identifying which types of students respond to such outreach activities. We find robust evidence of heterogeneous treatment effects by altruistic behaviour: students who report stronger altruistic preferences become significantly more likely to express interest in studying Economics after the talk.

Given that the average treatment effect is close to zero, the positive response among more altruistic students implies a weaker—or, in some specifications, slightly negative—response among less altruistic students. This suggests that the intervention primarily operates through a compositional shift in expressed interest rather than an overall expansion of demand for Economics. While this pattern does not necessarily represent a drawback, it highlights that outreach activities may influence who is attracted to the discipline more than how many students are attracted in total. If one views the under-representation of more altruistic students in Economics as a concern (as a non-negligible body of literature implies), our results suggest that short, interactive popularisation talks may be a low-cost tool to move the pool of prospective Economics majors closer to the wider student population in terms of distributional preferences.

This pattern is robust to different estimation models, suggesting that it is not driven by functional form or sample selection. This heterogeneity effect aligns closely with mechanisms emphasised in previous experimental studies. Porter and Serra (2020) demonstrate that exposure to role models who challenge stereotypes about Economics can have a meaningful impact on students' major choices, particularly among groups whose preferences or identities are not well aligned with the discipline's traditional image. Our findings suggest that the way the intervention frames Economics as an empirical social science engaged with real-world problems may be especially salient for students with stronger distributional concerns, making them more willing to consider Economics as a university degree.

Taken together, the results suggest that popularisation efforts may have limited effects on average interest in the short run but can play a meaningful role in attracting students with altruistic motivations to Economics. This perspective helps reconcile small or imprecisely estimated average effects with the broader evidence that targeted informational and framing interventions can contribute to diversifying the pipeline into the economics profession (Bayer et al., 2020b).

Several limitations should be noted. First, our outcome captures stated interest—whether students include Economics among their top three preferred degrees—rather than actual subsequent enrolment in an Economics programme. The treatment may therefore affect short-run salience or expressed preferences without translating into realised choices. Relatedly, because the intervention was delivered face-to-face and immediately preceded the survey, part of the estimated effect may reflect an experimenter demand effect (Zizzo, 2010), in which students adjust their reported interest in response to perceived expectations rather than a durable change in educational plans. Our results should therefore be interpreted as capturing short-run interest in Economics immediately after the intervention, rather than durable changes in students' longer-term educational choices.

A related issue concerns the interpretation of the outcome variable. Students reported their preferred degree programmes from a list of courses offered by the host university. As a result, the measure may partly capture engagement with the host institution rather than interest in Economics as a discipline more generally. However, the content of the popularisation talk focused on the nature of Economics as a field of study and on its relevance for understanding social and policy problems, rather than on the university itself. This reduces the likelihood that the observed effects primarily reflect institutional marketing.

Second, altruistic preferences are measured using a self-reported item and may therefore be affected by social desirability bias. Although random assignment implies that such bias should be balanced across treatment and control groups, any resulting measurement error is likely to attenuate the estimated heterogeneous effects.

Third, the intervention is context-specific: the sample consists of high school students visiting a private university in one region of Spain. As a result, participants represent a selected group of students who are already sufficiently engaged with higher education to take part in university outreach activities, and a large share report having at least one parent with a university degree. The findings should therefore be interpreted as evidence from a specific outreach setting rather than as representative of the broader student population. At the same time, information frictions about Economics may be even greater among students with less exposure to higher education, suggesting that the effects of similar informational interventions could differ in other contexts.

## 6. Conclusion

This paper examines whether a short popularisation talk about Economics affects high school students' interest in studying the subject. We find no statistically significant average treatment effect, although the point estimates imply a non-negligible increase in stated interest. More importantly, we uncover robust heterogeneity by altruistic preferences: students with stronger altruistic preferences become significantly more likely to report interest in studying Economics after the intervention. This pattern is consistent across linear and nonlinear specifications and remains after addressing missing data through multiple imputation.

These findings suggest that short popularisation talks about Economics may have limited effects on average interest in the short run, but they can influence who is attracted to Economics by disproportionately engaging students whose motivations align with the discipline's social and policy-oriented dimensions. In other words, the intervention appears to reallocate interest towards students with stronger prosocial motivations rather than increasing aggregate interest across the board. From this perspective, such interventions may play a role in shaping the composition of prospective Economics students, even when average effects are modest. In particular, students with altruistic preferences respond when Economics is presented as empirical and socially relevant, concerned with inequality and social problems, rather than as a purely technical discipline. Therefore, departments interested in attracting such students can adjust their messaging to highlight the social science and problem-solving dimensions of Economics. If sustained, such compositional changes may have downstream consequences for the discipline's culture and research priorities. Our findings also suggest that outreach efforts emphasising an empirical and socially relevant view of Economics may help Economics attract students who might otherwise self-select out of the field.

Overall, our results suggest that low-cost popularisation talks about Economics, while not generating a statistically significant increase in average interest in studying the subject, can nonetheless meaningfully shape *who* becomes interested in it. More specifically, students with stronger altruistic (distributional) preferences respond positively to an intervention that presents Economics as an empirical social science engaged with real-world problems and that involves direct personal interaction. From a methodological perspective, these findings highlight that outreach efforts should not be evaluated solely on the basis of aggregate effects: even when average impacts are modest or insignificant, such interventions can alter the composition of the pool of students who express interest in Economics. For universities and economics departments, this suggests that targeted communication strategies may be an effective tool for attracting students with specific motivations and values, potentially contributing over time to a more socially oriented and diverse discipline.

## Appendix A. Presentation used in the popularisation talk and survey form

We include both the original Spanish version and an English translation of the presentation used in the talks and the survey form. The presentation is exactly the original one used in the popularisation talks. For the survey form, we removed all information, logos and colours that could identify the university or the authors.

# ¿QUÉ HACEN LOS ECONOMISTAS?

Nombre y afiliación de los autores

## ¿QUÉ HACEN LOS ECONOMISTAS?

- Qué es la Medicina y cuál es el trabajo de un médico…
- No ocurre con otras profesiones y disciplinas, como, por ejemplo…
- …la Economía
- **¿QUÉ ES?**
- **¿DE QUÉ SE OCUPA?**

## ¿QUÉ HACEN LOS ECONOMISTAS?

- Jacob Viner:
- ***"Economía es lo que hacen los economistas"…***
- Objetivo de la charla = explicaros qué es la Economía a partir de algunos ejemplos reales en su quehacer diario

# ¿A QUÉ ASOCIAS LA ECONOMÍA?

## La creencia popular

Mercados financieros

Tipos de interés

Bancos

Criptomonedas

## Las preguntas habituales

¿En qué invierto mi dinero?

¿Es buen momento para comprar una casa?

¿Qué es eso del bitcoin?

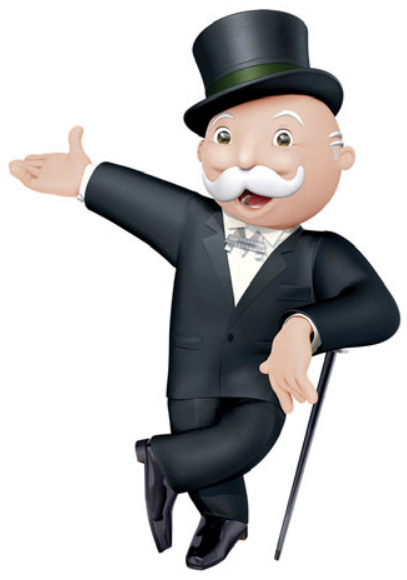

## ¿QUÉ HACEN LOS ECONOMISTAS?

- Muchos economistas trabajan en bancos…
- Realizan inversiones en bolsa o …
- Son capaces de dar consejos sobre qué hacer con nuestros ahorros
  - Ojo…
- Identificar y reducir la Economía a estos clichés…
- …es equivocado…

## LA ECONOMÍA ES UNA DISCIPLINA MUCHO, MUCHO MÁS AMPLIA…

- Ciencia **Social**
  - Como Sociología, Psicología, Ciencia Política…
  - Estudia el **comportamiento humano**
- Estudio de cómo las **personas interactúan** entre sí y con su entorno natural para obtener sus medios de subsistencia, y cómo esto cambia con el tiempo

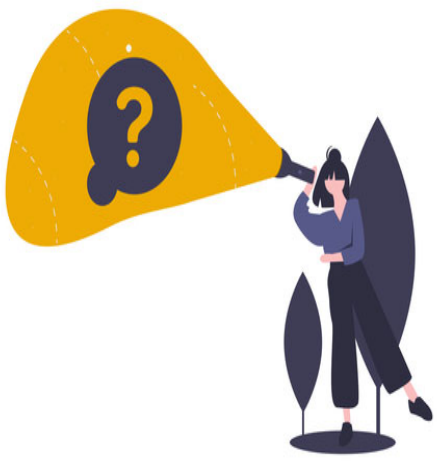

## LA ECONOMÍA ES UNA DISCIPLINA MUCHO, MUCHO MÁS AMPLIA…

- La Economía elabora **teorías**…
- …basadas en **modelos matemáticos**…
  - Simplificaciones de la realidad
- y luego tratar de ver si esas teorías son correctas o no
  - Con métodos estadísticos

## ¡ESTUDIA PROBLEMAS <u>SOCIALES</u>!

- ? Por qué unos países son ricos y otros no
- ? El empleo y el desempleo
- ? La pobreza y la desigualdad
- ? El cambio climático
- ? Educación
- ? Salud

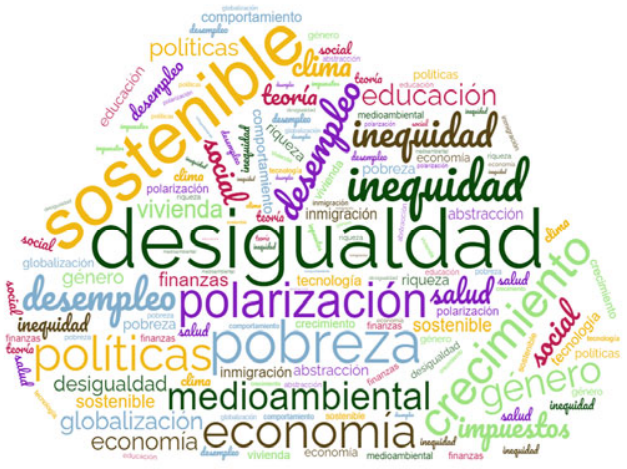

## ¡ES UNA CIENCIA!

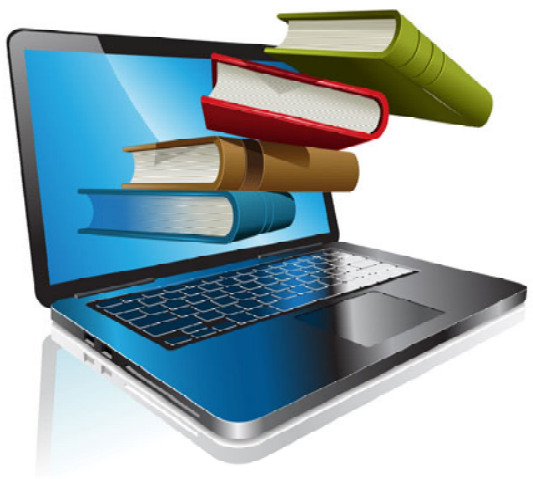

- Propone teorías que deben ser verificadas o rechazadas en la realidad
  - Si rechazadas, nuevas teorías….
- Utiliza el **método científico** (proceso ordenado para explicar fenómenos reales; problema, observación, hipótesis, predicción, experimentación, conclusión): es **aplicada** y se basa en la **observación** y **medición**

## ¡ES UNA CIENCIA!

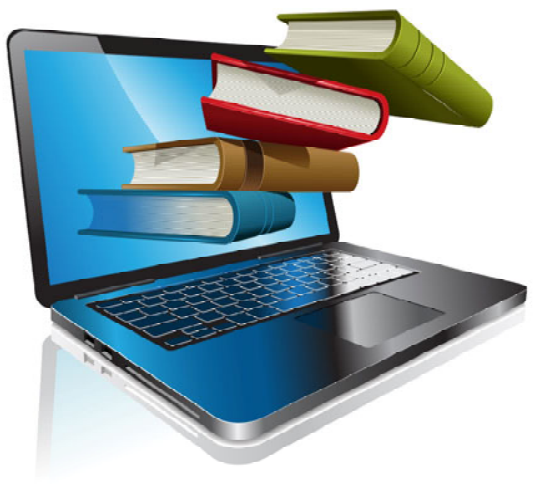

- Para observar y medir, emplea **experimentos** o técnicas que proporcionan resultados muy similares a los experimentos (sorteo control, tratamiento)

# EDUCACIÓN

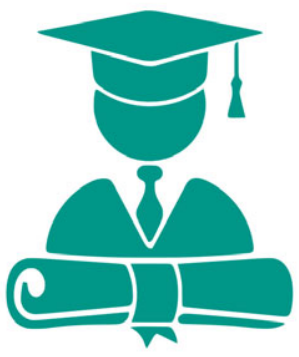

Aplicación 1: EDUCACIÓN

## ¿CUÁL ES EL EFECTO DE REDUCIR EL NÚMERO DE ESTUDIANTES EN LAS CLASES?

Experimentos EEUU; cuasi experimentos en Suecia (leyes reducen tamaño clases). Estudiar en clases más pequeñas mejora el rendimiento académico y aumentan los salarios de los estudiantes.

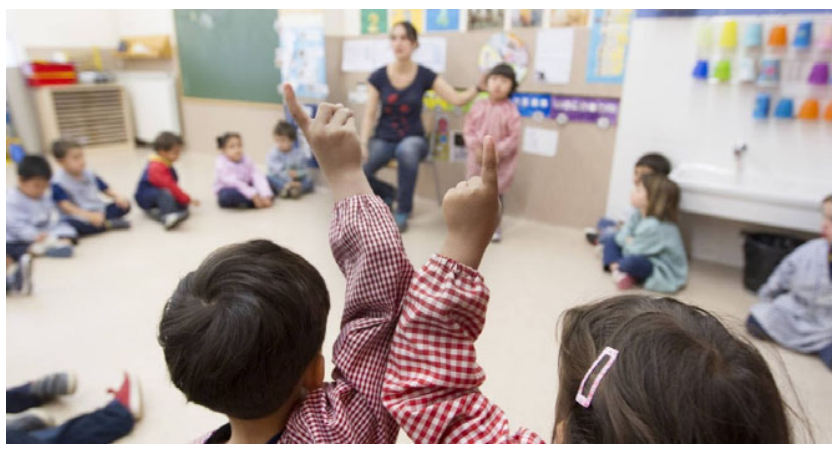

# SALUD

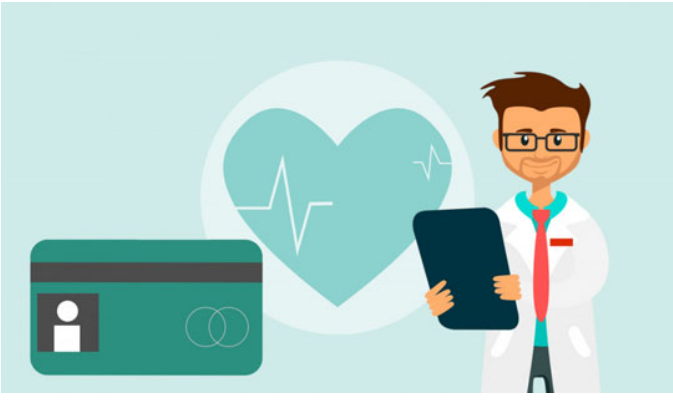

Aplicación 2: SALUD

## ¿SIRVEN LAS MASCARILLAS PARA PROTEGERNOS FRENTE AL COVID-19?

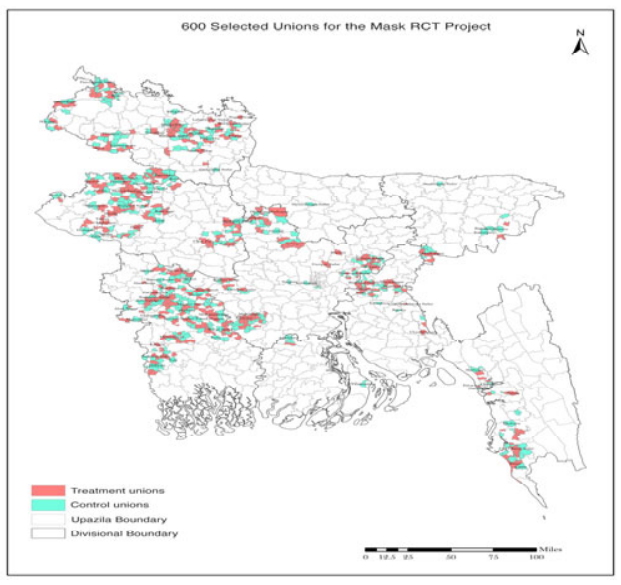

Fuente: Abaluck et al. (2021). Impact of community masking on COVID-19: A cluster-randomized trial in Bangladesh. *Science*, *375*(6577), 160.

Un experimento en Bangladesh (300 pueblos mascarilla obligatoria vs. 300 no obligatoria, aleatoriamente) mostró que el uso de mascarillas redujo los contagios el 10%.

Aplicación 2: SALUD

## ¿CUÁL ES EL EFECTO DE PROHIBIR LOS MÓVILES DURANTE LA CONDUCCIÓN?

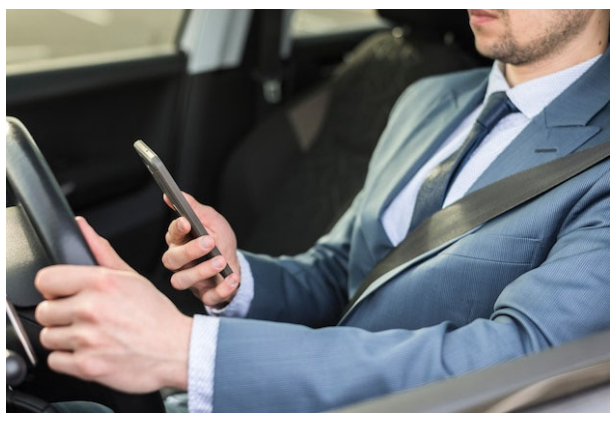

Imagen de <a href="https://www.freepik.es/foto-gratis/hombre-negocios-smartphone-coche_3386231.htm#query=creative%20commons%20coche%20tel%C3%A9fono&position=12&from_view=search&track=ais">Freepik</a>

En Estados Unidos, cada estado puede establecer sus propias normas sobre uso móvil. La prohibición salvó numerosas vidas humanas al reducir los accidentes (69 vidas por región/año)

# DISCRIMINACIÓN

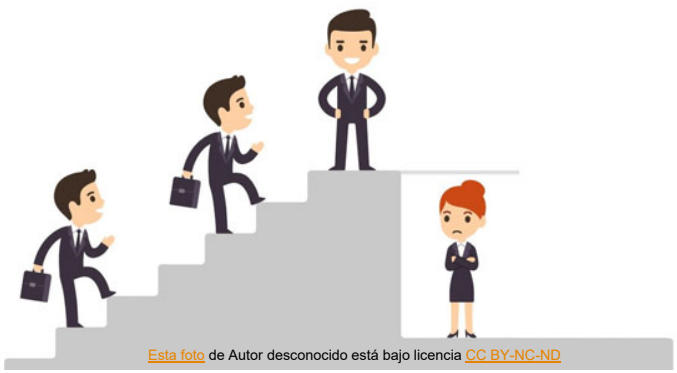

Aplicación 4: DISCRIMINACIÓN

## ¿ESTÁN DISCRIMINADAS LAS TRABAJADORAS INMIGRANTES PROCEDENTES DE PAÍSES MUSULMANES?

Se manda CV falsos a ofertas de trabajo. Única diferencia: foto y nombre. Con el mismo nivel educativo y cualificaciones, tener un nombre turco y llevar hiyab (velo en la cabeza) reducía las posibilidades de empleo.

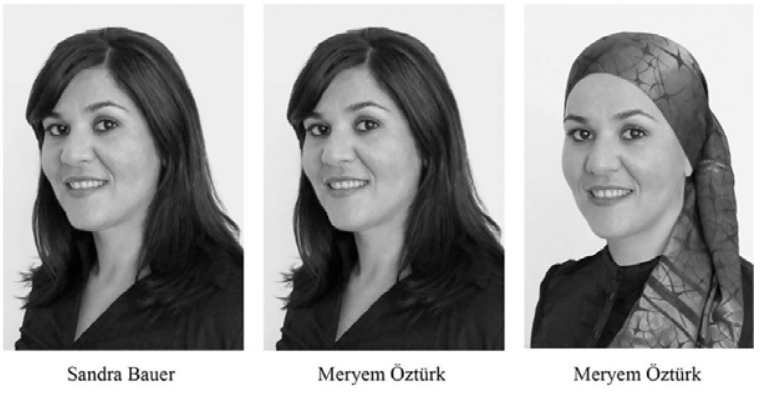

Fuente: Weichselbaumer, D. (2020). Multiple discrimination against female immigrants wearing headcarves. *ILR Review*, *73*(3), 600-627.

## ¿DÓNDE PODEMOS ENCONTRAR ECONOMISTAS?

- Bancos e instituciones financieras
- Empresas consultoras
- Investigación en centros públicos o privados
- Profesores en institutos o universidades
- En empresas privadas (informes coyuntura, previsiones económicas, análisis diferentes tecnologías)
- En empresas públicas (análisis impacto políticas, fondos europeos, programas educativos, de salud…)
- Economistas del Estado

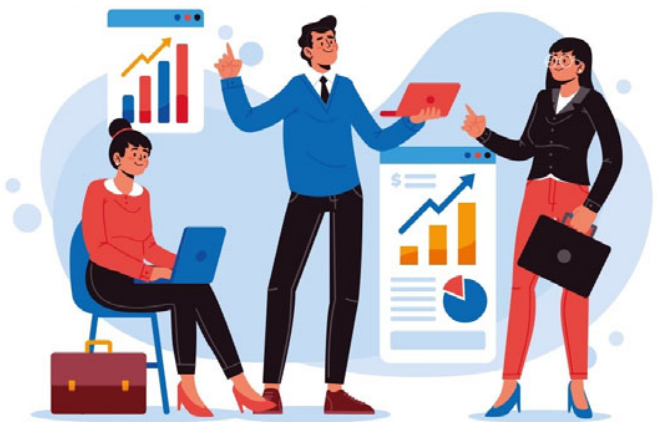

## ¿DÓNDE PODEMOS ENCONTRAR ECONOMISTAS?

- En ONG
- En organismos internacionales (ONU, UE, OCDE, Banco Mundial, FMI, BID…)
- Organismos reguladores de la competencia
- Servicio de estudios (análisis mercado laboral…)
- Empresas de seguros (evaluación riesgos y costes de diferentes coberturas…)

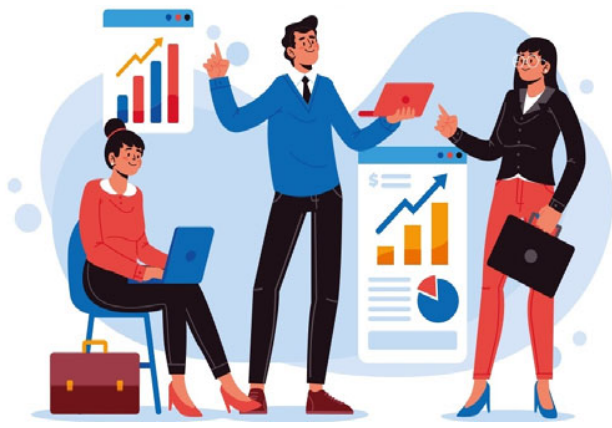

# WHAT DO ECONOMISTS DO?

Authors' names and affiliations

## WHAT DO ECONOMISTS DO?

- What is Medicine and what does a doctor do?
  - Everyone has a general idea
- This is not the case with other professions and disciplines, such as...
- ...Economics
- **WHAT IS IT?**
- **WHAT DOES IT DEAL WITH?**

## WHAT DO ECONOMISTS DO?

- Jacob Viner:
- ***"Economics is what economists do"...***
- Aim of the talk = to explain what economics is, using some real examples from their daily work.

# WHAT DO YOU ASSOCIATE ECONOMICS WITH?

## Popular belief

Financial markets

Interest rates

Banks

Cryptocurrencies

## Frequently asked questions

Where should I invest my money?

Is it a good time to buy a house?

What is bitcoin?

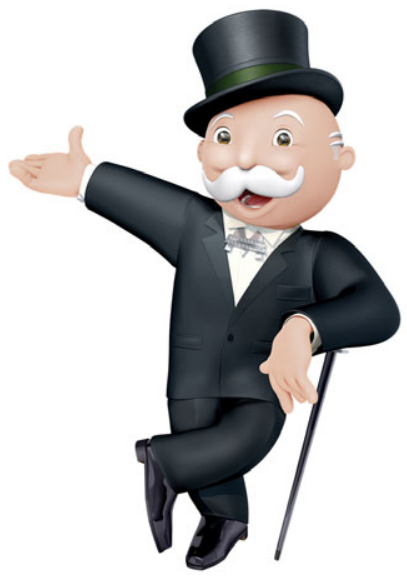

## WHAT DO ECONOMISTS DO?

- Many economists work in banks...
- They make investments on the stock market or...
- They are able to give advice on what to do with our savings
  - Be careful...
- Identifying and reducing economics to these clichés...
- ...is wrong...

## ECONOMICS IS A MUCH, MUCH BROADER DISCIPLINE...

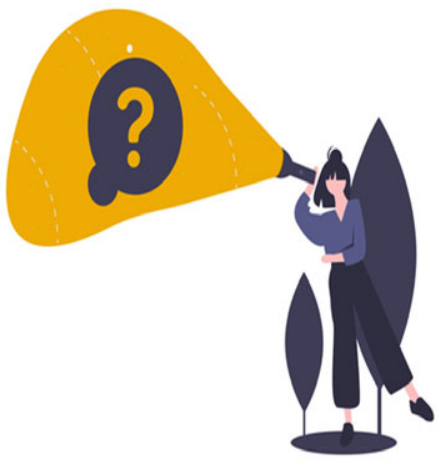

- **Social** Science
  - Like Sociology, Psychology, Political Science...
  - Studies **human behaviour**
- The study of how **people interact** with each other and their natural environment to obtain their livelihood, and how this changes over time

## ECONOMICS IS A MUCH, MUCH BROADER DISCIPLINE...

- Economics develops theories...
- **...**based on mathematical **models**...
  - Simplifications of reality
- and then tries to see if those theories are correct or not
  - Using statistical methods

## IT STUDIES SOCIAL PROBLEMS!

- Why some countries are rich and other are not
- Employment and unemployment
- Poverty and inequality
- Climate change
- Education
- Health

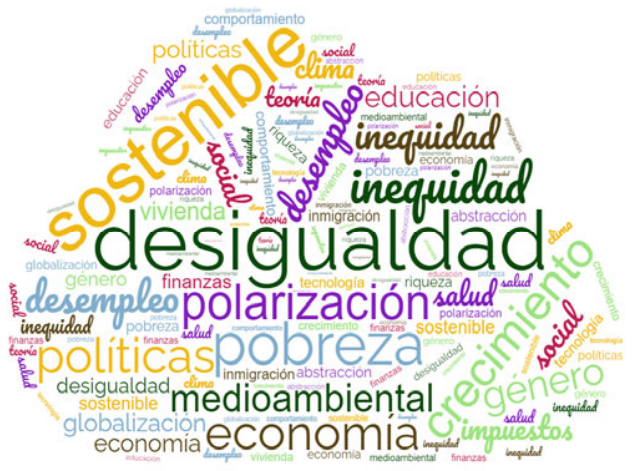

## IT IS A SCIENCE!

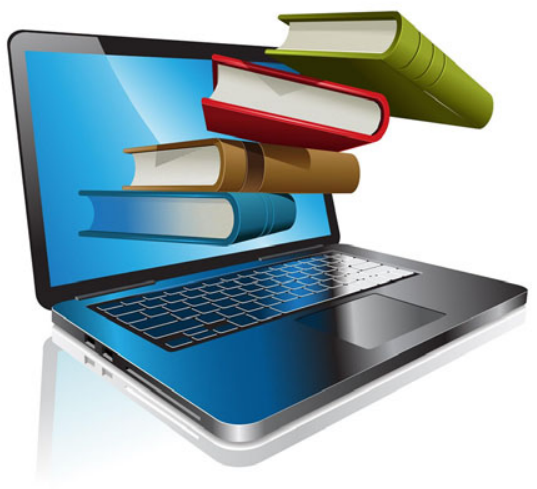

- It proposes theories that must be verified or rejected in reality.
- If rejected, new theories...
- It uses the **scientific method** (an orderly process for explaining real phenomena: problem, observation, hypothesis, prediction, experimentation, conclusion): it is **applied** and based on **observation** and **measurement**

## IT IS A SCIENCE!

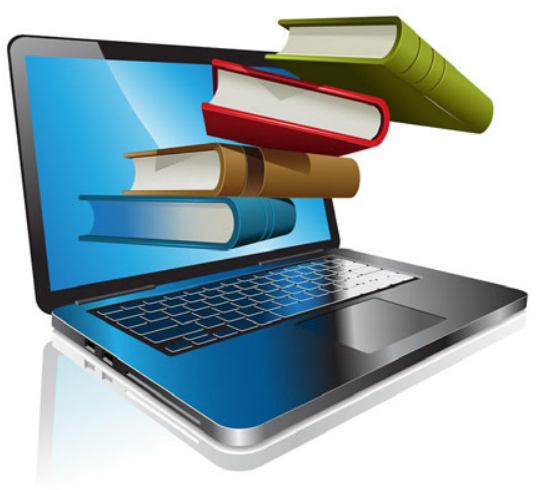

- To observe and measure, it uses **experiments** or techniques that provide results very similar to experiments (controlled randomisation, treatment).

# EDUCATION

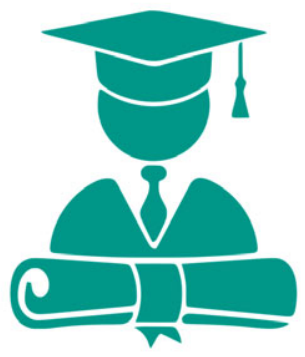

**Application 1: EDUCATION**

## WHAT IS THE EFFECT OF REDUCING THE NUMBER OF STUDENTS IN CLASSES?

Experiments in the US; quasi-experiments in Sweden (laws reducing class sizes). Studying in smaller classes improves academic performance and increases students' salaries.

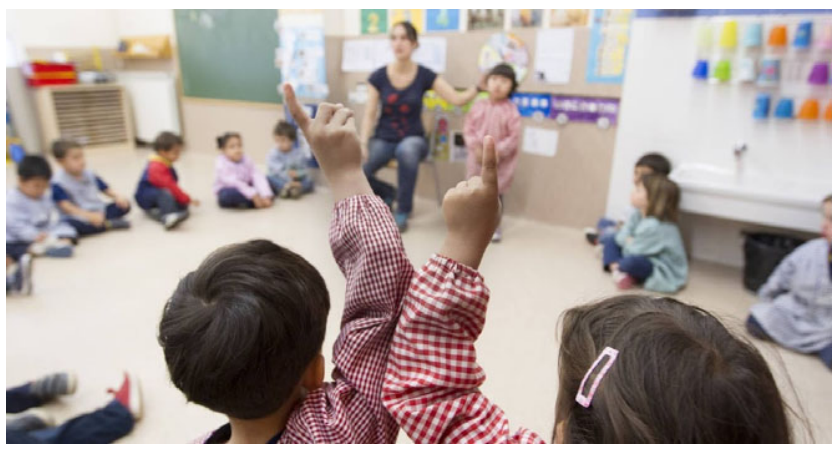

Picture by Unknown author licensed under license CC BY-SA-NC

# HEALTH

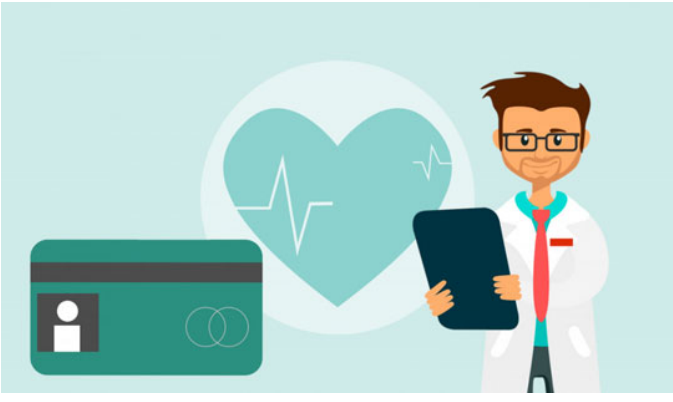

**Application 2: HEALTH**

## DO MASKS PROTECT US AGAINST COVID-19?

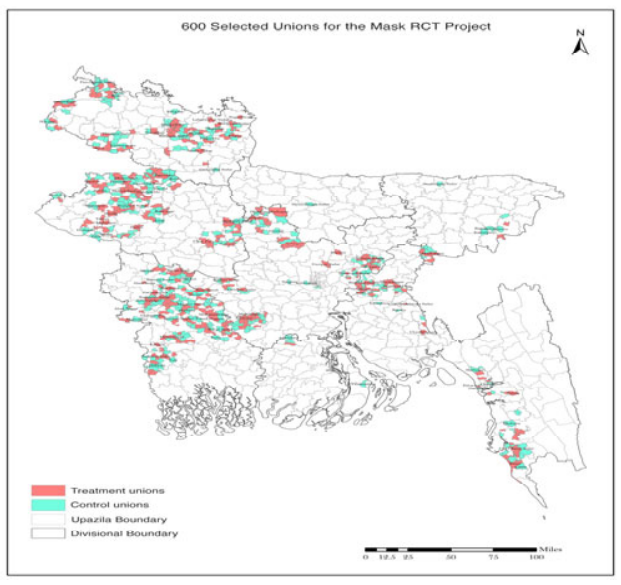

Source: Abaluck et al. (2021). Impact of community masking on COVID-19: A cluster-randomized trial in Bangladesh. *Science*, *375*(6577), 160.

An experiment in Bangladesh (300 villages with mandatory masks vs. 300 without, randomly selected) showed that wearing masks reduced infections by 10%.

**Application 3: HEALTH**

## WHAT IS THE EFFECT OF BANNING MOBILE PHONES WHILE DRIVING?

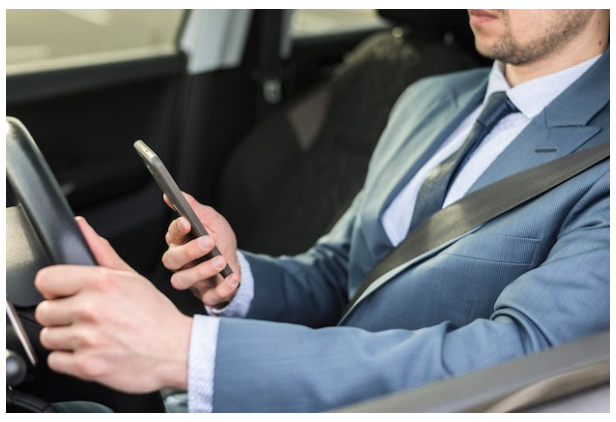

Image from <a href="https://www.freepik.es/foto-gratis/hombre-negocios-smartphone-coche_3386231.htm#query=creative%20commons%20coche%20tel%C3%A9fono&position=12&from_view=search&track=ais">Freepik</a>

In the United States, each state can establish its own rules on mobile phone use. The ban saved numerous lives by reducing accidents (69 lives per region/year).

# DISCRIMINATION

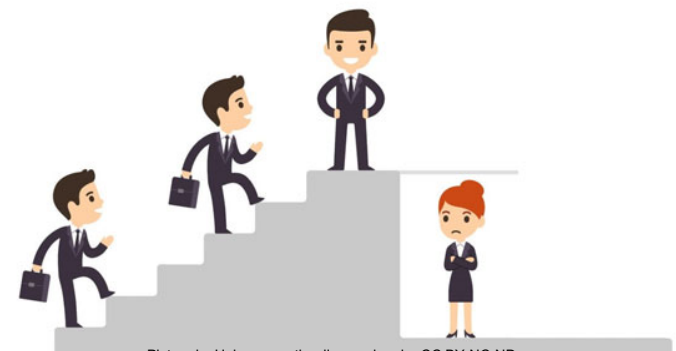

Picture by Unknown author licensed under CC BY-NC-ND.

**Application 4: DISCRIMINATION**

## ARE IMMIGRANT WORKERS FROM MUSLIM COUNTRIES DISCRIMINATED AGAINST?

Fake CVs are sent to job offers. The only difference: photo and name. With the same level of education and qualifications, having a Turkish name and wearing a hijab (headscarf) reduced the chances of employment.

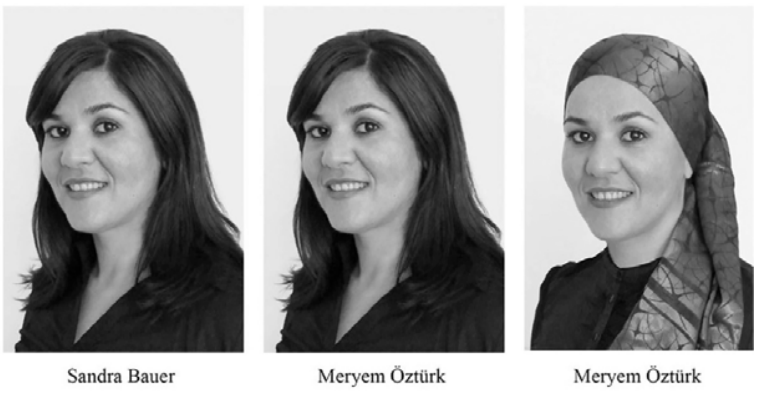

Source: Weichselbaumer, D. (2020). Multiple discrimination against female immigrants wearing headcarves. *ILR Review*, 73(3), 600-627.

## WHERE CAN WE FIND ECONOMISTS?

- ✓ Banks and financial institutions
- ✓ Consulting firms
- ✓ Research in public or private centres
- ✓ Teachers in secondary schools or universities
- ✓ In private companies (economic reports, economi forecasts, analysis of different technologies)
- ✓ In public companies (policy impact analysis, European funds, educational and health programr etc.)
- ✓ State economists

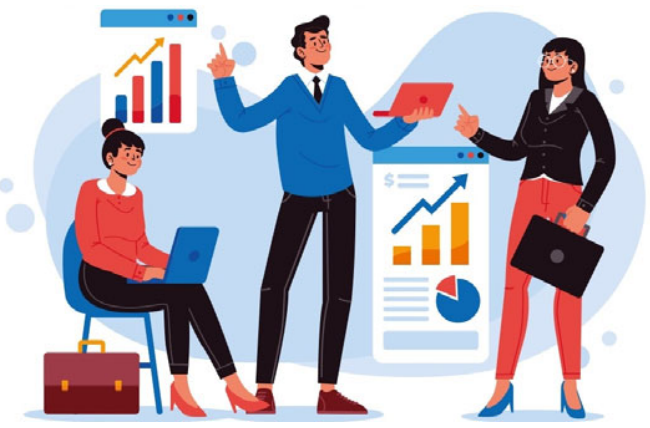

## WHERE CAN WE FIND ECONOMISTS?

- ✓ In NGOs
- ✓ In international organisations (UN, EU, OECD, World Bank, IMF, IDB, etc.)
- ✓ Competition regulatory bodies
- ✓ Research departments (labour market analysis, etc.)
- ✓ Insurance companies (risk assessment and costs of different types of cover, etc.)

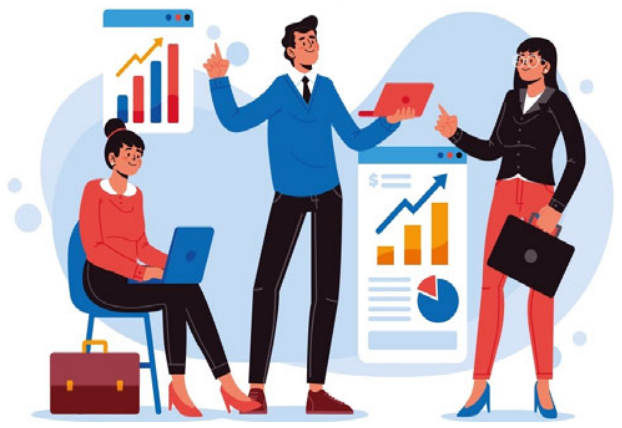

# SOLICITUD DE INFORMACIÓN

Logo de la Universidad 1

**DATOS PERSONALES, CENTRO Y ESTUDIOS ACTUALES** (Rellenar con mayúscula)

NOMBRE Y APELLIDOS

E-MAIL

MÓVIL

POBLACIÓN

NOMBRE DEL CENTRO

MODALIDAD DE BACHILLERATO | CURSO ACTUAL | NOTA MEDIA

## GRADOS

**1** ADE Bilingüe **A**B
**2** ADE **A**B
**3** Artes Escénicas y Cinematográficas **A**
**4** Biotecnología **A**
**5** Ciencias de la Actividad Física y del Deporte **A**B
**6** Ciencias Religiosas **C**
**7** Comunicación Audiovisual **A**B
**8** Creación de Videojuegos y Animación Digital **A**
**9** Creación y Producción Musical **A**
**10** Criminología **A**B
**11** Datos y Analítica de Negocio **A**B
**12** Derecho **A**B
**13** Economía **A**
**14** Educación Infantil **A**
**15** Educación Primaria **A**
**16** Enfermería **A**B
**17** Farmacia **A**
**18** Fisioterapia **A**
**19** Ingeniería de las Tecnologías Industriales **A**
**20** Ingeniería de Organización Industrial **A**
**21** Ingeniería del Software **A**
**22** Ingeniería Informática y de Tecnologías Virtuales **A**
**23** Ingeniería Mecatrónica y Robótica **A**
**24** Matemática Aplicada **A**
**25** Medicina **A**
**26** Nutrición Humana y Dietética **A**
**27** Periodismo y Medios Digitales **A**B
**28** Psicología **A**B
**29** Publicidad y Marketing Digital **A**B
**30** Relaciones Internacionales **A**B
**31** Teología **C**

## GRADOS DOBLES

**32** ADE + Datos y Analítica de Negocio **A**B
**33** ADE + Derecho **A**B
**34** ADE + Economía **A**
**35** ADE + Publicidad y Marketing Digital **A**B
**36** ADE + Relaciones Internacionales **A**
**37** ADE Bilingüe + Datos y Analítica de Negocio **A**B
**38** ADE Bilingüe + Derecho **A**B
**39** ADE Bilingüe + Publicidad y Marketing Digital **A**B
**40** ADE Bilingüe + Relaciones Internacionales **A**B
**41** Artes Escénicas y Cinematográficas + Creación y Producción Musical **A**
**42** Comunicación Audiovisual + Artes Escénicas y Cinematográficas **A**
**43** Comunicación Audiovisual + Creación y Producción Musical **A**
**44** Derecho + Criminología **A**
**45** Derecho + Relaciones Internacionales **A**B
**46** Economia + Datos y Analítica de Negocio **A**
**47** Educación Primaria + Educación Infantil **A**
**48** Farmacia + Nutrición Humana y Dietética **A**
**49** Ingeniería de las Tecnologías Industriales + ADE Bilingüe **A**
**50** Periodismo y Medios Digitales + Comunicación Audiovisual **A**B
**51** Periodismo y Medios Digitales + Publicidad y Marketing Digital **A**B
**52** Periodismo y Medios Digitales + Relaciones Internacionales **A**B
**53** Psicología + Criminología **A**B

**A** Campus A **B** Campus B **C** Campus C

**De los grados ofertados, indica, poniendo el NÚMERO correspondiente al grado, CUÁL SERÍA TU PRIMERA OPCIÓN (GRADO PREFERIDO), SEGUNDA Y TERCERA OPCIÓN.**

**GRADO PRIMERA OPCIÓN** | **GRADO SEGUNDA OPCIÓN** | **GRADO TERCERA OPCIÓN**

**¿Quieres solicitar una cita individual para informarte?** SÍ NO

**¿Qué titulación estudiarías que no esté en la lista anterior?**

**¿Consideras estudiar en la Universidad Loyola?** SÍ NO NS/NC

**¿Alguno de tus padres ha completado estudios universitarios?** SÍ NO

**Campus donde cursar los estudios:** A B C

**¿Con qué opción estás más de acuerdo?:**

No me importa el dinero que tenga; lo que me preocupa es que haya gente que tenga menos dinero que yo

No me importa el dinero que tenga; lo que me preocupa es que haya gente que tenga más dinero que yo

**Esta encuesta forma también parte de un proyecto de investigación sobre vocaciones profesionales de investigado-res de la Universidad 1 y de la Universidad 2. Has sido seleccionado como parte de una muestra para estudiar las vocaciones profesionales entre los estudiantes de secundaria. Nos gustaría hacerte unas breves preguntas relacionadas con estudios universitarios. La información recopilada para el estudio será tratada de forma agregada y confidencial, pero tus respuestas contribuirán al entendimiento de las vocaciones profesionales. Puedes firmar si deseas participar en el estudio. Para cualquier duda, contactar con: datos de contacto de uno de los autores.**

FIRMA

# REQUEST FOR INFORMATION

Logo of University 1

## PERSONAL DETAILS, SCHOOL AND CURRENT STUDIES (Write in capital letters)

FULL NAME

E-MAIL

MOBILE PHONE

CITY/TOWN

NAME OF SCHOOL

TYPE OF UPPER SECONDARY PROGRAMME | CURRENT YEAR | AVERAGE GRADE

## DEGREE PROGRAMMES

**1** Business Administration (Bilingual) **AB**
**2** Business Administration **AB**
**3** Performing and Film Arts **A**
**4** Biotechnology **A**
**5** Physical Activity and Sport Sciences **AB**
**6** Religious Studies **C**
**7** Audiovisual Communication **AB**
**8** Videogame Creation and Digital Animation **A**
**9** Music Creation and Production **A**
**10** Criminology **AB**
**11** Business Data and Analytics **AB**
**12** Law **AB**
**13** Economics **A**
**14** Early Childhood Education **A**
**15** Early Childhood Education **A**
**16** Nursing **AB**
**17** Pharmacy **A**
**18** Physioteraphy **A**
**19** Industrial Technologies Engineering **A**
**20** Industrial Organization Engineering **A**
**21** Software Engineering **A**
**22** Computer Engineering and Virtual Tecnologies **A**
**23** Mechatronics and Robotions Engineering **A**
**24** Applied Mathematics **A**
**25** Medicine **A**
**26** Human Nutrition and Dietetics **A**
**27** Journalism and Digital Media **AB**
**28** Psychology **AB**
**29** Digital Advertising and Marketing **AB**
**30** International Relations **AB**
**31** Theology **C**

## DOUBLE DEGREE PROGRAMMES

**32** Business Administration + Business Data Analytics **AB**
**33** Business Administration + Law **AB**
**34** Business + Economics **A**
**35** Business Administration + Digital Advertising and Marketing **AB**
**36** Business Administration + International Relations **A**
**37** Business Administration (Bilingual) + Business Data and Analytics **AB**
**38** Business Administration (Bilingual) + Law **AB**
**39** Business Administration (Bilingual) + Digital Advertising and Marketing **AB**
**40** Business Administration (Bilingual) +International Relations **AB**
**41** Performing and Film Arts + Music Creation and Production **A**
**42** Audiovisual Communication + Performing and Film Arts **A**
**43** Audiovisual Communication + Music Creation and Production **A**
**44** Law + Criminology **A**
**45** Law + International Relations **AB**
**46** Economics+ Business Data and Analytics **A**
**47** Primary Education + Early Childhood Education **A**
**48** Pharmacy + Human Nutrition and Dietetics **A**
**49** Industrial Technologies Engineering + Business Administration (Bilingual) **A**
**50** Journalism and Digital Media + Audiovisual Communication **AB**
**51** Journalism and Digital Media + Digital Advertising and Marketing **AB**
**52** Journalism and Digital Media + International Relations **AB**
**53** Psychology + Criminology **AB**

**A** Campus A **B** Campus B **C** Campus C

---

**From the degrees offered , indicate, by writing the corresponding NUMBER correspondiente al grado, WHICH WOULD BE YOUR FIRST CHOICE (PREFERRED DEGREE), SECOND CHOICE AND THIRD CHOICE.**

**FIRST CHOICE DEGREE** **SECOND CHOICE DEGREE** **THIRD CHOICE DEGREE**

---

**Would you like to request and individual information appointment?** YES NO

**What degree would you study that is not on the above list?**

**Are you considering studying at University 1?** YES NO DK/NA

**Have any of your parents completed university studies?** YES NO

**Campus where you would study:** A B C

**Which statement do yhou agree with more?:**

I do not care about how much money I have; what concerns me is that there are people who have less money than I do.

I do not care about how much money I have; what concerns me is that there are people who have more money than I do.

---

**This survey is also part of a research project on professional and academic aspirations conducted by researchers from University 1 and University 2. You have been selected as part of a sample to study vocational aspirations among secondary school students. We would like to ask you a few brief questions related to university studies. The information collected will be treated in an aggregated and confidential manner and your responses will contribute to a better understanding of professional aspirations. You may sign if you wish to participate in the study.For anyh queries, please contact: contact details of one of the authors.**

**In accordance with the General Data Protection Regulation (GDPR) of 27 April 2016, and in order to ensure the proper functioning of the University, we inform you that, should you have authorized the processing of personal data by selecting the relevant boxes on the admissions platform, you may revoke such authorization at any time, as indicated below. You may also exercise your rights of access, rectification, erasure, restriction, data portabilit, and objection under the terms established by law, by contacting the data controller: contact details of University 1**

SIGNATURE

# Appendix B: Additional analyses

Table B1. Attrition tests

| | (I) | (II) | (III) | (IV) | (V) | (VI) |
|---|---|---|---|---|---|---|
| | Panel A<br>Differential attrition rates | | | Panel B<br>Attrition conditional on observables | | |
| | Non-response (altruistic) | Non-response (GPA) | Non-response (altruistic & GPA) | Non-response (altruistic) | Non-response (GPA) | Non-response (altruistic & GPA) |
| Treatment | −0.029 | 0.107* | 0.043 | −0.004 | 0.094 | 0.052 |
| | (0.053) | (0.062) | (0.061) | (0.052) | (0.067) | (0.071) |
| | [0.581] | [0.115] | [0.496] | [0.942] | [0.275] | [0.508] |
| Female | | | | 0.064* | 0.028 | 0.076** |
| | | | | (0.034) | (0.027) | (0.035) |
| Parents with a university degree | | | | 0.018 | −0.014 | 0.005 |
| | | | | (0.046) | (0.041) | (0.052) |
| Senior | | | | 0.118*** | −0.254*** | −0.098 |
| | | | | (0.041) | (0.073) | (0.065) |
| Charter school | | | | −0.054 | 0.027 | −0.045 |
| | | | | (0.058) | (0.050) | (0.069) |
| Public school | | | | −0.089 | 0.061 | −0.041 |
| | | | | (0.057) | (0.083) | (0.086) |
| Observations | 1,313 | 1,313 | 1,313 | 1,211 | 1,211 | 1,211 |
| $R^2$ | 0.001 | 0.018 | 0.002 | 0.017 | 0.076 | 0.015 |
| Joint significance test of covariates | | | | | | |
| *F*-statistic | | | | 2.358 | 3.159 | 2.095 |
| *WCB p*-value | | | | 0.232 | 0.206 | 0.245 |

*Notes:* Columns (I), (II) and (III) report OLS regressions of each non-response dummy variable (and their combination) on treatment status. Columns (IV), (V) and (VI) add covariates to analyse attrition conditional on observables. Standard errors clustered at the school level (26 clusters) are reported in parentheses. WCB *p*-values are reported in brackets to account for the small number of clusters. Asterisks denote statistical significance based on cluster-robust standard errors: *** $p < 0.01$, ** $p < 0.05$, * $p < 0.10$. The last two rows report, for each regression, the *F*-statistic testing the joint null hypothesis that all covariate coefficients are equal to zero, along with the corresponding WCB *p*-value.

Table B2. Balance in the non-missing sample across treatment and control groups

| | (I) | (II) | (III) | (IV) | (V) | (VI) | (VII) | (VIII) | (IX) |
|---|---|---|---|---|---|---|---|---|---|
| | Panel A Non-missing data for altruistic | | | Panel B Non-missing data for GPA | | | Panel C Non-missing data for altruistic and GPA | | |
| | Control | Treatment | Difference (II) – (I) | Control | Treatment | Difference (V) – (IV) | Control | Treatment | Difference (VIII) – (VII) |
| Female | 0.544 | 0.475 | −0.070 | 0.544 | 0.484 | −0.060 | 0.534 | 0.456 | −0.078 |
| | (0.499) | (0.500) | (0.054) | (0.498) | (0.500) | (0.058) | (0.499) | (0.499) | (0.059) |
| | | | [0.228] | | | [0.336] | | | [0.234] |
| Parents with a university degree | 0.860 | 0.742 | −0.117 | 0.877 | 0.747 | −0.130* | 0.863 | 0.749 | −0.113 |
| | (0.348) | (0.438) | (0.070) | (0.329) | (0.435) | (0.069) | (0.345) | (0.434) | (0.070) |
| | | | [0.116] | | | [0.078] | | | [0.133] |
| Senior | 0.599 | 0.625 | 0.026 | 0.916 | 0.901 | −0.015 | 0.911 | 0.879 | −0.032 |
| | (0.320) | (0.376) | (0.134) | (0.278) | (0.299) | (0.089) | (0.285) | (0.327) | (0.103) |
| | | | [0.694] | | | [0.873] | | | [0.767] |
| Charter school | 0.884 | 0.830 | −0.055 | 0.477 | 0.317 | −0.160 | 0.516 | 0.300 | −0.217 |
| | (0.500) | (0.460) | (0.225) | (0.500) | (0.466) | (0.231) | (0.500) | (0.459) | (0.234) |
| | | | [0.406] | | | [0.516] | | | [0.398] |
| Public school | 0.503 | 0.303 | −0.200 | 0.032 | 0.334 | 0.303* | 0.035 | 0.350 | 0.315* |
| | (0.210) | (0.479) | (0.161) | (0.175) | (0.472) | (0.159) | (0.183) | (0.478) | (0.162) |
| | | | [0.406] | | | [0.094] | | | [0.086] |
| Altruistic | 0.046 | 0.354 | 0.307* | | | | 0.588 | 0.626 | 0.038 |
| | (0.491) | (0.485) | (0.040) | | | | (0.493) | (0.485) | (0.044) |
| | | | [0.538] | | | | | | [0.408] |
| GPA | | | | 8.598 | 8.463 | −0.135 | 8.570 | 8.413 | −0.157 |
| | | | | (1.204) | (1.203) | (0.250) | (1.163) | (1.230) | (0.249) |
| | | | | | | [0.621] | | | [0.555] |
| Joint orthogonality test | | | | | | | | | |
| *F*-statistic | | | $F(6,25) = 2.813$ | | | $F(6,25) = 2.171$ | | | $F(6,25) = 6.733$ |
| *p*-value | | | 0.393 | | | 0.493 | | | 0.184 |
| No. of observations | 519 | 376 | 895 | 665 | 404 | 1,069 | 461 | 297 | 758 |

*Notes:* The table reports mean values by treatment status and the difference between them, computed from OLS regressions of each covariate on the treatment indicator. Standard errors clustered at the school level (26 clusters) are used for statistical inference. WCB *p*-values are reported in brackets to account for the small number of clusters. Asterisks denote statistical significance based on cluster-robust standard errors: *** $p < 0.01$, ** $p < 0.05$, * $p < 0.10$. The joint orthogonality test follows Kerwin et al. (2024) and is based on randomisation inference. WCB is not used for this joint test, as the procedure relies on the exact random assignment mechanism rather than asymptotic approximations and remains valid with a limited number of clusters.

Table B3. Average treatment effects of the popularisation talk on students' interest in Economics (*logit* estimates)

| | (I) | (II) | (III) |
|---|---|---|---|
| Treatment | 0.022 | 0.021 | 0.004 |
| | (0.025) | (0.030) | (0.027) |
| Female | | −0.059*** | −0.049** |
| | | (0.023) | (0.025) |
| Parents with a university degree | | −0.028 | −0.022 |
| | | (0.019) | (0.030) |
| Senior | | 0.025 | 0.004 |
| | | (0.019) | (0.039) |
| Charter school | | 0.019 | 0.002 |
| | | (0.024) | (0.024) |
| Public school | | 0.007 | 0.017 |
| | | (0.033) | (0.038) |
| GPA | | | 0.000 |
| | | | (0.007) |
| Altruistic | | | −0.001 |
| | | | (0.020) |
| | | | |
| No. of observations | 1,313 | 1,211 | 730 |
| Pseudo R2 | 0.003 | 0.034 | 0.031 |
| Log likelihood | −347.5 | −312.1 | −161 |
| Correctly classified (%) | 92.54 | 92.49 | 93.97 |

*Notes:* Each column reports the average marginal effects obtained from *logit* estimates of the effect of the popularisation talk on students' interest in studying Economics under different sets of control variables. All specifications include an intercept. Standard errors clustered at the school level (26 clusters) are reported in parentheses. *** $p < 0.01$; ** $p < 0.05$; * $p < 0.10$.

Table B4. Effect heterogeneity by gender, parental education and distributional preferences (*logit* estimates)

| | (I) | (II) | (III) | (IV) | (V) | (VI) | (VII) | (VIII) | (IX) |
|---|---|---|---|---|---|---|---|---|---|
| | Panel A Heterogeneity by gender (H2) | | | Panel B Heterogeneity by parental education (H3) | | | Panel C Heterogeneity by distributional preferences (H4) | | |
| Treatment effect for males | −0.002 | −0.004 | −0.026 | | | | | | |
| | (0.037) | (0.042) | (0.046) | | | | | | |
| Treatment effect for females | 0.038 | 0.044 | 0.035 | | | | | | |
| | (0.025) | (0.028) | (0.022) | | | | | | |
| Treatment effect for students with parents with a university degree | | | | −0.030 | −0.025 | −0.039 | | | |
| | | | | (0.038) | (0.047) | (0.070) | | | |
| Treatment effect for students with parents without a university degree | | | | 0.035 | 0.030 | 0.013 | | | |
| | | | | (0.029) | (0.029) | (0.022) | | | |
| Treatment effect for students without altruistic preferences | | | | | | | −0.054* | −0.052 | −0.068** |
| | | | | | | | (0.032) | (0.037) | (0.034) |
| Treatment effect for students with altruistic preferences | | | | | | | 0.066*** | 0.068** | 0.056** |
| | | | | | | | (0.025) | (0.031) | (0.026) |
| No. of observations | 1,305 | 1,211 | 730 | 1,219 | 1,211 | 730 | 895 | 860 | 730 |
| Pseudo-$R^2$ | 0.030 | 0.041 | 0.042 | 0.011 | 0.037 | 0.035 | 0.0335 | 0.064 | 0.070 |
| Log likelihood | −337.6 | −309.8 | −159.2 | −320.0 | −311.1 | −160.5 | −210.2 | −193.8 | −154.7 |
| Correctly classified (%) | 92.49 | 92.49 | 93.97 | 92.53 | 92.49 | 93.97 | 93.41 | 93.49 | 93.97 |

*Notes:* Each column reports the average marginal effects obtained from *logit* estimates of the effect of the popularisation talk on students' interest in studying Economics, including the corresponding interaction term, under different sets of control variables. All specifications include an intercept; specifications in columns (I), (IV) and (VII) include no additional controls; columns (II), (V) and (VIII) add type of school and senior status; columns (III), (VI) and (IX) additionally include GPA. Standard errors clustered at the school level (26 clusters) are reported in parentheses. *** $p < 0.01$; ** $p < 0.05$; * $p < 0.10$.

Table B5. Average treatment effects of the popularisation talk on students' interest in Economics (OLS estimates using multiple imputation)

| | (I) | (II) | (III) | (IV) |
|---|---|---|---|---|
| Treatment | 0.021 | −0.003 | −0.028 | −0.035 |
| | (0.031) | (0.041) | (0.047) | (0.039) |
| Female | −0.056** | −0.075** | −0.055** | −0.057*** |
| | (0.021) | (0.033) | (0.020) | (0.020) |
| Parents with a university degree | −0.026 | −0.024 | −0.055* | −0.024 |
| | (0.020) | (0.020) | (0.028) | (0.020) |
| Senior | 0.026 | 0.027 | 0.026 | 0.023 |
| | (0.019) | (0.020) | (0.019) | (0.019) |
| Charter school | 0.015 | 0.015 | 0.012 | 0.012 |
| | (0.024) | (0.024) | (0.023) | (0.023) |
| Public school | 0.005 | 0.006 | 0.010 | 0.003 |
| | (0.036) | (0.036) | (0.035) | (0.037) |
| Altruistic | 0.002 | 0.001 | 0.002 | −0.034 |
| | (0.019) | (0.019) | (0.019) | (0.023) |
| GPA | −0.004 | −0.003 | −0.004 | −0.004 |
| | (0.007) | (0.007) | (0.007) | (0.007) |
| Treatment × female | | 0.046 | | |
| | | (0.038) | | |
| Treatment × parents with a university degree | | | 0.058 | |
| | | | (0.045) | |
| Treatment × altruistic | | | | 0.090** |
| | | | | (0.036) |
| No. of observations | 1,211 | 1,211 | 1,211 | 1,211 |
| $R^2$ | | | | |
| Treatment + interaction | | 0.042 | 0.030 | 0.055* |
| | | (0.031) | (0.031) | (0.032) |

Notes: Each column reports OLS estimates of the effect of the popularisation talk on students' interest in studying Economics using multiply imputed data. Missing values in altruistic preferences and GPA are imputed using multiple imputation by chained equations. All specifications include an intercept and control for gender, parental education, senior status, type of school, distributional preferences and GPA. Columns (II), (III) and (IV) include interactions between the treatment variable and gender, parental education and distributional preferences, respectively. The row labelled *treatment + interaction* reports the total effect for the interacted category. Standard errors clustered at the school level (26 clusters) are reported in parentheses. *** $p < 0.01$; ** $p < 0.05$; * $p < 0.10$.